\documentclass[
reprint,
superscriptaddress,
nofootinbib,
amsmath,amssymb,
aps,
pre,
]{revtex4-2}

\usepackage{graphicx}
\graphicspath{{Figures/serif/}}

\usepackage{dcolumn}
\usepackage{etoolbox,siunitx}
\robustify\bfseries
\usepackage{bm}
\usepackage{hyperref}
\usepackage{xfp} 
\usepackage{booktabs}

\usepackage[capitalise]{cleveref} 

\hypersetup{colorlinks=true, 
linkcolor=blue,
filecolor=blue,      
urlcolor=blue,
citecolor=blue,
}

\newcommand\Tstrut{\rule{0pt}{2.4ex}}       
\newcolumntype{C}[1]{>{\centering\arraybackslash}p{#1}}

\usepackage{upgreek}
\usepackage{bm}
\newcommand{\vect}[1]{\boldsymbol{\mathbf{#1}}}

\DeclareMathOperator*{\argmax}{arg\,max}

\newcommand{\ENz}{z} 
\newcommand{\ENe}{e} 
\newcommand{\ENou}{ou} 
\newcommand{\ENere}{re} 
\newcommand{\ENll}{ll} 
\newcommand{\ENLA}{\@} 

\newcommand{\ENhyphNon}{-} 

\usepackage[UKenglish]{babel}
\usepackage[UKenglish]{isodate}

\newcommand{\ENone}{\num{1}} 
\newcommand{\ENtwo}{\num{2}} 

\newcommand{\rsDateTrainStart}{1959-10-02}
\newcommand{\rsDateTrainEnd}{2008-09-01}
\newcommand{\rsDateDataEnd}{2020-11-20}

\newcommand{\rsQuantileBoth}{0.95}
\newcommand{\rsQuantileLeft}{\fpeval{(1 - \rsQuantileBoth)/2}}
\newcommand{\rsQuantileRight}{\fpeval{(1 + \rsQuantileBoth)/2}}

\newcommand{\rsThreshBoth}{0.01856}
\newcommand{\rsThreshLeft}{-0.01840}
\newcommand{\rsThreshRight}{+0.01872}

\newcommand{\rsMdlBi}{_\mathrm{bi}}
\newcommand{\rsMdlUni}{_\mathrm{bi}^{d}}
\newcommand{\rsMdlAci}{_\mathrm{ci}}
\newcommand{\rsMdlSci}{_\mathrm{ci}^{s}}

\newcommand{\rsTL}{\leftarrow}
\newcommand{\rsTR}{\rightarrow}
\newcommand{\rsTA}{\leftrightarrow}
\newcommand{\rsTO}{\leftrightharpoons}

\newcommand{\rsUdt}{\mathrm{d}_{\mathrm{t}}}

\newcommand{\mdUoBathAdd}{University of Bath, Bath BA2 7AY, United Kingdom}

\begin{document}

\preprint{APS/123-QED}

\title{Asymmetric excitation of left- and right-tail extreme events probed using a Hawkes model: application to financial returns
}

\author{Matthew F. Tomlinson}
	\email{mft28@bath.ac.uk}
\affiliation{Department of Physics, \mdUoBathAdd}
\affiliation{Centre for Networks and Collective Behaviour, \mdUoBathAdd}

\author{David Greenwood}
\affiliation{
CheckRisk LLP, 4 Miles's Buildings, George Street, Bath BA1 2QS, United Kingdom}

\author{Marcin Mucha-Kruczy\'{n}ski}%
\affiliation{Department of Physics, \mdUoBathAdd}
\affiliation{Centre for Nanoscience and Nanotechnology, \mdUoBathAdd}

\date{\today}

\begin{abstract}

We construct a two-tailed peaks-over-threshold Hawkes model that captures asymmetric self- and cross-excitation in and between left- and right-tail extreme values within a time series. We demonstrate its applicability by investigating extreme gains and losses within the daily log-returns of the S\&P 500 equity index. We find that the arrivals of extreme losses and gains are described by a common conditional intensity to which losses contribute twice as much as gains. However, the contribution of the former decays almost five times more quickly than that of the latter. We attribute these asymmetries to the different reactions of market traders to extreme upward and downward movements of asset prices: an example of negativity bias, wherein trauma is more salient than euphoria.

\end{abstract}

\maketitle

\section{\label{sec:intro}Introduction}

Heuristics such as imitation and herding are significant drivers of human agents within social systems. These reflexive behavi\ENou{}rs of individuals lead to self-exciting dynamics at the group level that often feature time-clustering of extreme events at the macro scale \cite{Kim2020, Reinhart2018}. Such extreme events often have profound consequences, which motivates a strong interest in their accurate forecasting. This problem is often approached through extreme value analysis (EVA), where asymptotic tail behavi\ENou{}r is mode\ENll{}ed independently from bulk behavi\ENou{}r, with the justification that the two are often generated by distinct mechanisms and, therefore, that the bulk provides little information about the tail and vice versa \cite{Coles2001, DeHaan2006, Scarrott2012}. Non\ENhyphNon{}stationary EVA methods that account for the time-clustering of extremes promise both improved forecasting accuracy and potential insight into the underlying mechanisms that generate extreme events. 

Peaks-over-threshold (POT) Hawkes models provide a parsimonious framework to describe macroscopic self-excitement of extreme values \cite{Hawkes1971a, Hawkes1971b}. In these models, the arrivals of threshold exceeding values within a time series become the discrete events of an inhomogeneous point process in which past events cause a time-decaying increase in the arrival rate of future events \cite{Reinhart2018, Hawkes2018}. Having first emerged as a stochastic model for the self-reflexive pattern of foreshocks and aftershocks that decorate major seismic activity \cite{Hawkes1971a, Hawkes1971b, Adamopoulos1976, Shcherbakov2019}, Hawkes-type models have since found application to broader classes of systems that exhibit similar activity bursts, including neural networks \cite{Pernice2012, Tannenbaum2017}, inter-group conflict \cite{Short2014, Johnson2018}, social media \cite{Fujita2018}, and financial markets \cite{Bacry2015, Hawkes2018, Hawkes2020, Bowsher2007, Bacry2012, Filimonov2012, Hardiman2013, Hardiman2014, Rambaldi2015, Grothe2014, Gresnigt2015, Gresnigt2016, Gresnigt2017, Chavez-Demoulin2005, Chavez-Demoulin2012, BienBarkowska2020}.

Here, we present a novel two-tailed peaks-over-threshold (2T-POT) Hawkes model that captures asymmetric self- and cross-excitation in and between left- and right-tail extreme values within the same univariate time series. This model assumes a conditional arrival intensity common to extreme events from both tails and is conceived as a stochastic model for the time clustering of extreme fluctuations within drift-diffusion-like processes. Previous work has seldom investigated the possibility of asymmetric interactions between the two sets of extremes; however, such interactions are a distinct possibility, especially, for example, in socially driven processes, where the guiding heuristics of human agents may include different responses to the two sets of tail events \cite{Baumeister2001, Rozin2001}. To illustrate this point, we apply our model to the extreme gains and losses within the historic daily log-returns of the S\&P 500 equity index; we use likelihood-based inference methods to compare its performance against the limiting case of symmetric interactions between tails, as well as to a bivariate Hawkes model in which left- and right-tail extremes are treated as the events of two distinct point processes -- both with and without cross-excitement between them.

Financial asset price time series as a class represent an ideal case study for our model. Prices are often mode\ENll{}ed as geometric random walks, in which the log-returns (i.e.\ENLA{} changes in the log-price) are independent and identically distributed (i.i.d.\@) white noise \cite{Bachelier1900, Ruppert2015a}. However, contrary to this description, log-returns are characteri\ENz{}ed by heavy-tailed marginal distributions and positively autocorrelated conditional heteroscedasticity (a styli\ENz{}ed fact known as \textit{volatility clustering}) \cite{Cont2001, Davies2015, Ruppert2015a, Tsay2010}. Accordingly, extreme price fluctuations - measured as large magnitude log-returns - cluster in time, especially within periods of sustained overall negative price growth. These bursts of extremes are evident in \cref{fig:SP_N_delta_t_KS} for the S\&P 500 daily log-returns: 
\begin{figure*}
\includegraphics{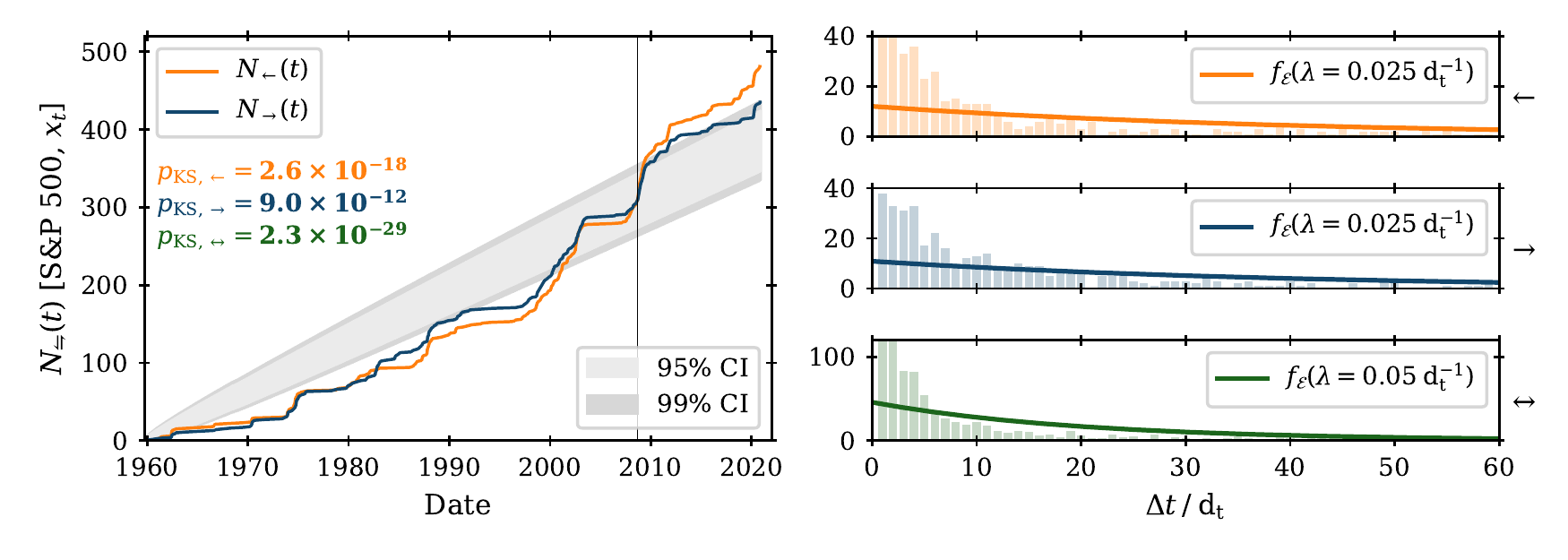}
\caption{
\label{fig:SP_N_delta_t_KS}
Arrival processes of extreme S\&P 500 daily log-returns. Left-tail (right-tail) extremes are defined as daily log-returns less (greater) than the 2.5\% (97.5\%) sample quantile in the training period. Left panel: left- (light orange) and right-tail (dark blue) exceedance count against time. The Kolmogorov-Smirnov (KS) test \cite{Lopes2011} is used against the null hypothesis implied by i.i.d.\@ log-returns, i.e.\ENLA{} $H_0: d{N_{\rsTO}}/d{t} \equiv \lambda_{\rsTO}{\left(t\right)} = 0.025 \: \rsUdt^{-1}$ (where $\rsUdt$ denotes trading days): the gr\ENe{}y shaded areas show the 95\% (lighter) and 99\% (darker) KS confidence intervals; also shown are the training period KS $p$-values for the left-tail ($N_{\rsTL}$), right-tail ($N_{\rsTR}$), and both-tails ($N_{\rsTA}$) processes, all of which are rejected at the 95\% confidence level. The vertical black line marks the end of the training period on \rsDateTrainEnd{}. Right panels: histogram of the interarrival times $\Delta{t}$ for exceedances from the left tail (top-right panel, light orange), right tail (middle-right panel, dark blue), and both tails (bottom-right panel, green); the solid lines show the expected exponential distribution under the assumption of i.i.d.\@ log-returns.
}
\end{figure*}
in the left panel, they manifest as step-like increases in the count of extreme returns as a function of time; in the right panels, we observe that short interarrival times between such extremes are much more frequent than expected under the null hypothesis of i.i.d.\@ returns. Our 2T-POT Hawkes approach is of particular interest here, because extreme gains and losses, while highly correlated \cite{Embrechts2011}, tend to be described by asymmetric distributions and persistence relationships \cite{Cont2001, Davies2015, Ruppert2015a, Tsay2010, Chicheportiche2014}. Moreover, heuristics such as negativity bias and loss aversion (i.e.\ENLA{} the tendency for human agents to prefer avoiding losses to acquiring equivalent gains) are well established within behavi\ENou{}ral economics \cite{Rabin1998, Khaneman2003}, and these could be expected to have effects at the group level, including asymmetric excitation of extremes. This is also supported by the leverage effect -- a styli\ENz{}ed fact of financial returns that states that the standard deviation of returns becomes larger when the average of returns becomes more negative \cite{Cont2001, Ruppert2015a}. Indeed, our model suggests that extreme losses contribute significantly more (by a factor of \ENtwo{}) than extreme gains to the conditional intensity. However, their importance as a function of time decays more rapidly.

We construct our Hawkes models in \cref{sec:model}. In \cref{sec:analysis}, we apply them to the daily log-returns of the S\&P 500 index between \rsDateTrainStart{} and \rsDateDataEnd{}; their performance is then evaluated through likelihood-based inference and residual analysis. In \cref{sec:CV}, we demonstrate that our observations cannot be wholly attributed to asymmetric volatility clustering as described by generali\ENz{}ed autoregressive conditional heteroscedasticity (GARCH) models, even after including the leverage effect. In \cref{sec:ML}, we detail our procedure of parameter estimation for the 2T-POT Hawkes model.

\section{\label{sec:model}Hawkes models for two-tailed threshold exceedances}

Starting from the discrete time series $\{x_{t}\}$, where $t$ indexes the data points (and is effectively the time as measured in trading days, a unit for which we use the symbol $\rsUdt$), we extract two sets of extreme events, $\{m_{k_{\rsTL}}\}=\{x_{t}-u_{\rsTL}<0\}$ and $\{m_{k_{\rsTR}}\}=\{x_{t}-u_{\rsTR}>0\}$, where $u_{\rsTL}$ and $u_{\rsTR}$ are the thresholds for the left and right tails of the data distribution, respectively, and $k_{\rsTL}$ and $k_{\rsTR}$ index the left- and right-tail exceedance events. Note that here we use the subscripts $\rsTL$ and $\rsTR$ to denote the left and right tail, respectively, and the subscript $\rsTO$ is used to represent either tail (i.e.\ENLA{} either $\rsTL$ or $\rsTR$) in generic expressions. We thus extract two point processes ${N_{\rsTO}}{\left(t\right)}$, wherein events are fully described by their arrival time $t_{k_{\rsTO}}$ and excess magnitude $m_{k_{\rsTO}}$, such that
\begin{equation}
\label{eqn:N}
d{N_{\rsTO}}{\left(t\right)} = \sum_{k_{\rsTO}}{\delta{\left(t-t_{k_{\rsTO}}\right)}} ,
\end{equation}
where $\delta{\left(t'\right)}$ is the Dirac delta function. The arrival rate of events within either point process is the conditional intensity for that process,
\begin{equation}
\label{eqn:lambda_defn}
\lambda_{\rsTO}{\left(t \middle| \mathcal{I}_t\right)} = \mathbb{E}{\left[ \frac{d{N_{\rsTO}}{\left(t\right)}}{dt} \middle| \mathcal{I}_t\right]},
\end{equation}
where $\mathbb{E}{[.]}$ is the expectation operator. The explicit time-dependence of $\lambda_{\rsTO}{\left(t \middle| \mathcal{I}_t\right)}$ specifies ${N_\rsTO}{\left(t\right)}$ to be \textit{inhomogeneous} point processes; Hawkes-type behavi\ENou{}r is specified by the conditional dependence on the event history up to the present time $t$, $\mathcal{I}_t = \left\{\left(t_{k_{\rsTO}}, m_{k_{\rsTO}}\right): t_{k_{\rsTO}} < t\right\}$.

\subsection{\label{ssec:models_bi}Bivariate 2T-POT Hawkes model}

\begin{figure*}
\includegraphics{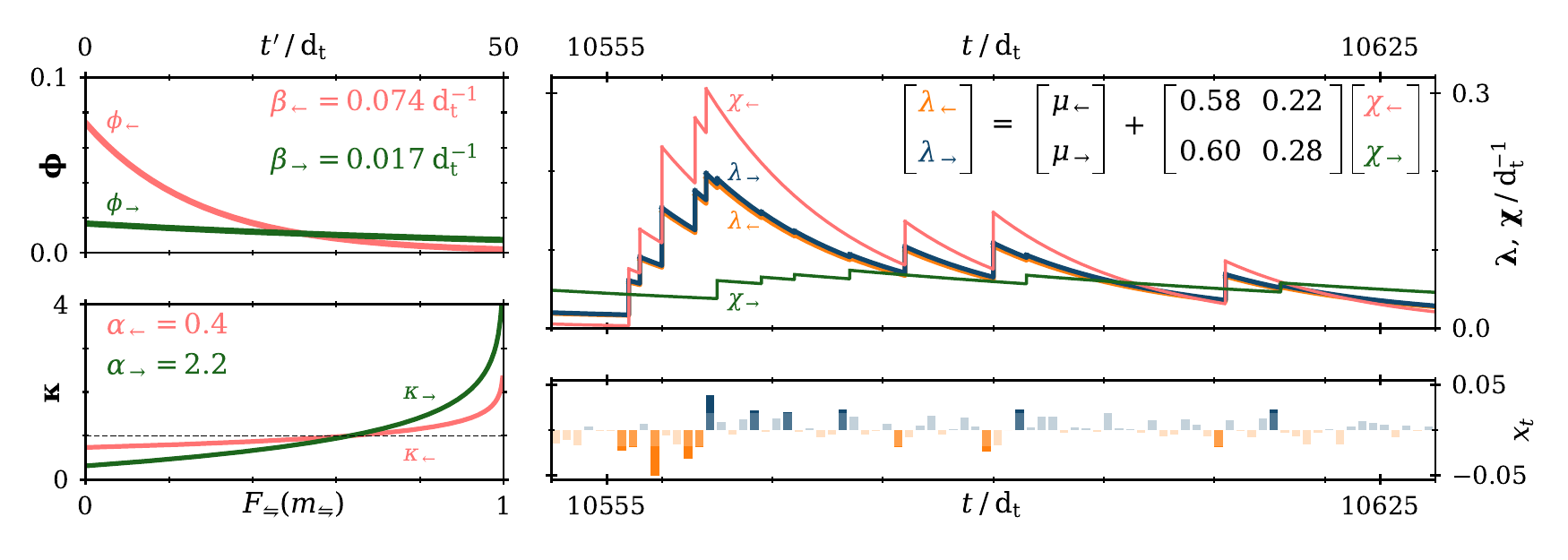}
\caption{
\label{fig:SP_process_bi-mark}
Bivariate 2T-POT Hawkes model $\vect{\uptheta}_{\rsMdlBi}$ fitted to historic S\&P 500 daily log-returns. Left panels: endogenous excitement time kernels $\vect{\upphi}$ (top-left panel) and excess magnitude impact functions $\vect{\upkappa}$ (bottom-left panel) for left- (light red) and right-tail (dark green) exceedance events. Right panels: response (top-right panel, equation shows the fitted values of the branching matrix $\vect{\Gamma}$) of the conditional intensities $\vect{\uplambda}$ ($\lambda_{\rsTL}$ in light orange, $\lambda_{\rsTR}$ in dark blue; N.B.\@ $\lambda_{\rsTL}$ and $\lambda_{\rsTR}$ overlap almost perfectly because the rows of $\vect{\Gamma}$ are approximately equal) and endogenous excitements $\vect{\upchi}$ ($\chi_{\rsTL}$ in light red, $\chi_{\rsTR}$ in dark green) to a sample cluster of extreme S\&P 500 daily log-returns (bottom-right panel, excesses marked in bolder shading).
}
\end{figure*}

We start our discussion from the most general description: a bivariate Hawkes model in which the left- and right-tail exceedances are treated as distinct point processes. Note that this is incompatible with the fact that the arrivals of left- and right-tail events within the original time series are mutually exclusive. We remedy this in \cref{ssec:models_ci} and here only point out that the bivariate description becomes an increasingly valid approximation as $\lambda_{\rsTO} \to 0$. With this in mind, we write the bivariate model,
\begin{equation}
\label{eqn:bivariate_model_long}
\left(\begin{matrix}
			\lambda_{\rsTL} \\
			\lambda_{\rsTR} \\
\end{matrix}\right)
= 		
\left(\begin{matrix}
			\mu_{\rsTL} \\
			\mu_{\rsTR} \\
\end{matrix}\right)	
+
\left(\begin{matrix}
			\gamma_{\rsTL\rsTL} & \gamma_{\rsTL\rsTR}\\
			\gamma_{\rsTR\rsTL} & \gamma_{\rsTR\rsTR}\\
\end{matrix}\right)
\left(\begin{matrix}
			\chi_{\rsTL} \\
			\chi_{\rsTR} \\
\end{matrix}\right) , 
\end{equation}
or, in vector notation,
\begin{equation}
\label{eqn:bivariate_model}
\vect{\uplambda}{\left(t \middle| \vect{\uptheta}\rsMdlBi{}; \mathcal{I}_t\right)} = \vect{\upmu} + \vect{\Gamma} \vect{\upchi}{\left(t \middle| \vect{\uptheta}\rsMdlBi{}; \mathcal{I}_t\right)} ,
\end{equation}
where $\vect{\uptheta}\rsMdlBi{}$ is the parameter vector for the bivariate model, $\vect{\upmu} \equiv (\mu_{\rsTL}, \mu_{\rsTR})^T$ are the constant exogenous background intensities for each process, $\vect{\Gamma}$ is the $2 \times 2$ branching matrix, and $\vect{\upchi}$ are the endogenous excitements generated by the arrivals of events in each respective process. The response of this model, as parametri\ENz{}ed below, to a sample activity cluster is shown in \cref{fig:SP_process_bi-mark}.

The endogenous excitements $\vect{\upchi}$ are the sums of contributions from all past events within each process,
\begin{equation} \label{eqn:1DHawkesEndo}
\chi_{\rsTO}{\left(t \middle| \vect{\uptheta}; \mathcal{I}_t\right)} = \sum_{k_{\rsTO}:t_{k_{\rsTO}}<t}{\phi_{\rsTO}{\left(t-t_{k_{\rsTO}}\right)\kappa_{\rsTO}{\left(m_{k_{\rsTO}}\right)}}} ,
\end{equation} 
where the time kernel $\phi_{\rsTO}$ is a monotonically decreasing function of the time between the arrival of the past event, $t_{k_{\rsTO}}$, and the present, $t$. Here, this is taken as an exponential decay with constant $\beta_{\rsTO}$,
\begin{equation} \label{eqn:phi_exp}
\phi_{\rsTO}{\left(t'\right)} = \beta_{\rsTO}e^{-\beta_{\rsTO} t'} ,
\end{equation}
which allows for \cref{eqn:1DHawkesEndo} to be recast in Markov form,
\begin{equation} \label{eqn:1DHawkesEndo_Markov}
d{\chi_{\rsTO}} = \beta_{\rsTO} \left[- \chi_{\rsTO} d{t} + \kappa_{\rsTO}{\left(m_{k_{\rsTO}}\right)} d{N_{\rsTO}} \right] .
\end{equation}
The impact function $\kappa_{\rsTO}$ is a monotonically increasing function of the excess magnitude $m_{k_{\rsTO}}$. Following the approach of \cite{Grothe2014, Gresnigt2015}, this is defined so that the intensity jump from the exceedance event arriving at time $t_{k_{\rsTO}}$ is determined by the conditional quantile of $m_{k_{\rsTO}}$ at the arrival time,
\begin{equation} \label{eqn:kappa}
\kappa_{\rsTO}{\left(m_{k_{\rsTO}}\right)} = \frac{1 - \alpha_{\rsTO} \ln{\left[1 - {F_{{\rsTO}}}{\left(m_{k_{\rsTO}} \middle| t_{k_{\rsTO}}\right)}\right]}}{1 + \alpha_{\rsTO}},
\end{equation}
where $F_{{\rsTO}}$ is the cumulative distribution function for the excess magnitudes. When the mark parameter $\alpha_{\rsTO}>0$, larger magnitude events produce greater jumps in the excitement. Crucially, this reduces sensitivity on the choice of threshold value, since $\kappa_{\rsTO}{\left(m_{k_{\rsTO}}\right)} \rightarrow (1+\alpha_{\rsTO})^{-1}$ as $|m_{k_{\rsTO}}| \rightarrow 0$. Conversely, when $\alpha_{\rsTO}=0$, $\kappa_{\rsTO}$ becomes unity and we recover an \textit{unmarked} Hawkes process in which $\chi_{\rsTO}$ is independent of the magnitudes of past events. Also, note that $\mathbb{E}{\left[\kappa_{\rsTO}{\left(m\right)}\right]} \equiv 1$ for all values of $\alpha_{\rsTO}$. 

The excess magnitudes are assumed to be described by a conditional generali\ENz{}ed Pareto distribution (GPD). This choice is motivated by the Pickands-Balkema-de Haan theorem \cite{Pickands1975, Balkema1974}, which states that the GPD is the limiting distribution for linearly rescaled threshold excesses within a series of i.i.d.\@ random variables \footnote{The GPD has become the classical asymptotically motivated distribution for threshold excesses within extreme value analysis for this reason \cite{Scarrott2012}.}. Moreover, since the GPD is specified with a shape parameter $\xi_{\rsTO}$, it can describe a range of tail heaviness from faster-than-exponential decay ($\xi_{\rsTO} < 0$) to increasingly leptokurtic power-law decay ($\xi_{\rsTO} > 0$). The cumulative density function for excess magnitudes is
\begin{equation} \label{eqn:F_GPD}
{F_{\rsTO}}{\left(m \middle| t\right)} = 
\begin{cases}
	1 - {\left[1 \mp \xi_{\rsTO}\frac{m}{\sigma_{\rsTO}{\left(t\right)}}\right]}^{-1/\xi_{\rsTO}}, & \xi_\rsTO \neq 0, \\
	1 - \exp{\left[\pm m/\sigma_{\rsTO}{\left(t\right)}\right]}, & \xi_{\rsTO} = 0,  \\
\end{cases}
\end{equation}
where conditional dependence on the excess (i.e.\ENLA{} non\ENhyphNon{}background) intensity of the Hawkes process is introduced via the conditional scale parameter
\begin{equation} \label{eqn:sigma}
\sigma_{\rsTO}{\left(t\right)} = \varsigma_{\rsTO} + \eta_{\rsTO} \left[\lambda_{\rsTO}{\left(t\right)} - \mu_{\rsTO} \right] .
\end{equation}
Thus, when $\eta_{\rsTO} > 0$, larger magnitude events become more likely in high activity clusters, as is generally observed in price data \cite{Cont2001}. Conversely, when $\eta_{\rsTO} = 0$ the excess magnitudes are drawn from an unconditional GPD with scale parameter $\varsigma_{\rsTO}$.

The self-exciting dynamics of the Hawkes process can be understood as a branching process, in which \textit{daughter} events are triggered by the additional endogenous intensity produced by the arrival of prior \textit{mother} events. $\vect{\Gamma}$ is called the branching matrix, because $\gamma_{ij}$ is the mean number of daughter events in the process $N_i$ that are triggered by a mother event in the process $N_j$. This is so because the time kernel $\phi_{\rsTO}$ and impact function $\kappa_{\rsTO}$ are normali\ENz{}ed, such that the expected lifetime contribution of each event in process $N_{\rsTO}$ to the endogenous excitement $\chi_{\rsTO}$ is \ENone{}; this normali\ENz{}ation also guarantees that the model is uniquely fitted. The process is subcritical (i.e.\ENLA{} non\ENhyphNon{}explosive) provided the spectral radius of branching matrix $\rho{(\vect{\Gamma})}$ is less than \ENone{} \cite{Wheatley2019}. 

Overall, the bivariate model is characteri\ENz{}ed by a set of parameters, $\vect{\uptheta}\rsMdlBi{} =\{
	\vect{\upmu}, 
	\vect{\Gamma},
	\vect{\upbeta},
	\vect{\upxi},
	\vect{\varsigma},
	\vect{\upeta},
	\vect{\alpha}
\}$, where vector quantities are of the form, $\vect{\upmu} \equiv (\mu_{\rsTL}, \mu_{\rsTR})^T$. Note that the two distinct Hawkes processes describing each tail can be decoupled by applying the constraints $\gamma_{\rsTL\rsTR} = 0 = \gamma_{\rsTR\rsTL}$, i.e.\ENLA{} we recover two independent univariate Hawkes processes between which there is no cross-excitation. This decoupled bivariate 2T-POT Hawkes model is denoted by the parameter vector $\vect{\uptheta}\rsMdlUni{}$.

\subsection{\label{ssec:models_ci}Common intensity 2T-POT Hawkes model}

As noted in the beginning of \cref{ssec:models_bi}, the bivariate model assumes that the arrivals of left- and right-tail exceedances form two distinct point processes. This, however, does not guarantee that these two types of events are mutually exclusive. To enforce this requirement, we assume that both sets of exceedance arrivals constitute the events of a single, common point process, $N_{\rsTA}$, whose arrival rate is given by the one-dimensional common conditional intensity, $\lambda_{\rsTA}$. 

To develop a general common intensity model that allows for asymmetric cross-excitation between asymmetric tails, we modify \cref{eqn:bivariate_model} by reducing the branching matrix $\vect{\Gamma}$ to the branching vector $\vect{\upgamma}_{\rsTA}^{T} \equiv \left< \vect{1} \middle| \vect{\Gamma}\right. \equiv (\gamma_{\rsTA\rsTL}, \gamma_{\rsTA\rsTR})$ and by reducing the background intensity to a scalar, $\mu_{\rsTA} \equiv \left< \vect{1} \middle| \vect{\upmu}\right> \equiv \mu_{\rsTL} + \mu_{\rsTR}$. Thus,
\begin{equation}
\label{eqn:common_intensity_model}
{\lambda_{\rsTA}}{\left(t \middle| \vect{\uptheta}\rsMdlAci{}; \mathcal{I}_t\right)} = \mu_{\rsTA} + \vect{\upgamma}_{\rsTA}^{T} \vect{\upchi}{\left(t \middle| \vect{\uptheta}\rsMdlAci{}; \mathcal{I}_t\right)} ,
\end{equation}
where $\vect{\uptheta}\rsMdlAci{} =\{
	\mu_{\rsTA}, 
	\vect{\upgamma}_{\rsTA},
	\vect{\upbeta},
	\vect{\upxi},
	\vect{\varsigma},
	\vect{\upeta},
	\vect{\alpha},
	w_{\rsTA}
\}$ is the parameter vector for the common intensity model.

Each event is then stochastically drawn from either tail upon arrival just as the excess magnitude is also randomly sampled. This can be reali\ENz{}ed as the excess magnitude being drawn from a probability distribution that is a weighted piecewise union of the left- and right-tail distributions, i.e.\ENLA{} from a probability density function of the form
\begin{equation}
\label{eqn:common_intensity_m} 
f_{\rsTA}{\left(m\right)} = 
\begin{cases}
	S{\left(-w_{\rsTA}\right)} f_{\rsTL}{\left(m\right)} ,& m < 0,  \\
    S{\left(+w_{\rsTA}\right)} f_{\rsTR}{\left(m\right),} & m > 0,  \\
\end{cases}
\end{equation}
where $f_{\rsTO}$ are the probability density functions for the left- and right-tail excess magnitude distributions, and the weighting of probability between the two tails is determined by the logistic function,
\begin{equation}
\label{eqn:logistic} 
S{\left(w_{\rsTA}\right)} = 1/(1+e^{- w_{\rsTA}}) ,
\end{equation}   
with the tail-weight asymmetry parameter $w_{\rsTA}$, such that the relative frequency of left- to right-tail events is $\mathbb{E}{[N_{\rsTL}/N_{\rsTR}]} = \exp{\left(-w_{\rsTA}\right)}$.

Note that, if $w_{\rsTA}=0$ and all parameters in $\vect{\uptheta}\rsMdlAci{}$ are constrained to be symmetric (i.e.\ENLA{} so that the left- and right-tail components of all vector parameters are equal), then the common intensity 2T-POT model is equivalent to a single-tail POT Hawkes model applied to the absolute values of a copy of the original time series that is cent\ENere{}d on the mid-point between the thresholds, $x_{t}^* = x_{t} - (u_{\rsTR} + u_{\rsTL})/2$. That is, the set of absolute exceedances $\{|m_{k_{\rsTA}}|\}=\{|x_{t}^*| - u_{\rsTA} > 0\}$, where $u_{\rsTA} = (u_{\rsTR} - u_{\rsTL})/2$, is a union of $\{|m_{k_{\rsTL}}|\}$ and $\{|m_{k_{\rsTR}}|\}$, and a univariate Hawkes model applied to this exceedance series describes equal self- and cross-excitation between left and right tails that are symmetric in all properties. This symmetric common intensity 2T-POT model is denoted by the parameter vector $\vect{\uptheta}\rsMdlSci{}$.

\section{\label{sec:analysis}Application to S\&P 500 daily log-returns}

To demonstrate the utility of the two-tailed extension to the classic POT Hawkes model, we apply the 2T-POT Hawkes models developed in \cref{sec:model} to the daily log-returns of the S\&P 500 equity index between \rsDateTrainStart{} and \rsDateDataEnd{}. The data are partitioned into an in-sample training period and an out-of-sample forecast period, with the former ending (and the latter beginning) on \rsDateTrainEnd{}. The data were sourced from Yahoo Finance \cite{YahooFinance2020}.

In accordance with previous literature \cite{Chavez-Demoulin2005, Grothe2014, Gresnigt2015, Embrechts2011, BienBarkowska2020}, we do not attempt to normali\ENz{}e the local volatility before fitting the 2T-POT models and we use constant threshold values. This raises the question of whether extreme events should be defined with respect to the magnitude of recent behavi\ENou{}r or against the longer history of the time series. We choose the latter for practical considerations. First, model-based estimates of local volatility introduce assumptions to which residual extreme events are highly sensitive; thus, POT Hawkes models applied to the residual series tend to be dominated by the noise these assumptions introduce (see \cref{sec:CV}). Secondly, POT Hawkes models applied to the raw series of log-returns have been found to more accurately forecast tail behavi\ENou{}r than conditional volatility models \cite{BienBarkowska2020}; we intend to explore this aspect further in a follow-up paper. 

\begin{figure}[t]
\includegraphics{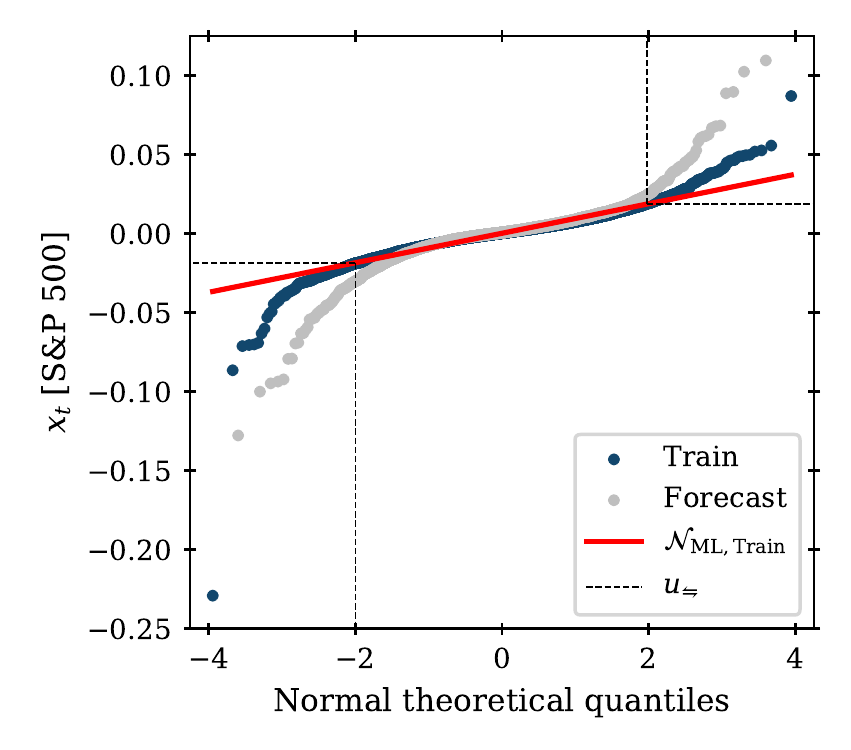}
\caption{
\label{fig:SP_x_qq}
Normal quantile plot of the S\&P 500 daily log-returns $x_t$ in the training (dark blue) and forecasting (light gr\ENe{}y) periods. The exceedance thresholds (black, dashed) coincide with the visual divergence of the training data from the maximum likelihood (ML) fitted normal distribution $\mathcal{N}_{\mathrm{ML, Train}}$ (red).  
}
\end{figure}

We acknowledge that, under our definition of extreme events, symmetric self- and cross-excitement may be perceived as an artefact of volatility clustering. However, one advantage of our two-tailed approach is that we may identify asymmetries that cannot be likewise attributed to volatility clustering as described by standard conditional volatility models. If, for instance, the log-returns are generated by a $\mathrm{GARCH}{(1,0,1)}$ process \cite{Ruppert2015a}, then, under mean-symmetric thresholds, we would expect all vector parameters of the fitted 2T-POT models to be symmetric, i.e.\ENLA{} $\hat{\vect{\uptheta}}\rsMdlSci{}$ would be the optimal model (this is explored, along with more complex cases, including the GJR-GARCH leverage effect \cite{Glosten1993}, in \cref{sec:CV}). To test this, we set the fixed threshold values according to a symmetric pair of sample quantiles within the training period. This also guarantees an equal number of left- and right-tail training period exceedances, and, therefore, that $w_{\rsTA} = 0$ for the common intensity models. Guided by where the marginal distribution of training period log-returns diverges from the normal distribution (\cref{fig:SP_x_qq}), we set the threshold values to the \fpeval{(100*\rsQuantileLeft)}\% and \fpeval{(100*\rsQuantileRight)}\% sample quantiles. We have verified that the results reported here are robust against small changes of these threshold values. Hereafter, we refer to exceedances of the two thresholds as \textit{extreme losses} and \textit{extreme gains}.

The parameters of each model are estimated from the training period data through the maximum likelihood (ML) procedure detailed in \cref{sec:ML}. The parameter estimates are listed in \cref{tab:params_mark_bi} for the bivariate models and in \cref{tab:params_mark_ci} for the common intensity models.

\begin{table*}
\caption{
L-BFGS-B parameter estimates ($\pm$ standard errors) for the coupled and decoupled bivariate 2T-POT Hawkes models trained on the extreme losses and gains of the S\&P 500 daily log-returns from \rsDateTrainStart{} to \rsDateTrainEnd{}.}
\label{tab:params_mark_bi}
\begin{ruledtabular}
\begin{tabular}{cllll}
\multicolumn{1}{c}{} & \multicolumn{2}{c}{$\hat{\vect{\uptheta}}\rsMdlBi{}$} &  \multicolumn{2}{c}{$\hat{\vect{\uptheta}}\rsMdlUni{}$}\\
\cmidrule{2-3}\cmidrule{4-5}
\multicolumn{1}{c}{Parameter} & \multicolumn{1}{c}{$\rsTL$\footnotemark[1]} &  \multicolumn{1}{c}{$\rsTR$\footnotemark[2]} & \multicolumn{1}{c}{$\rsTL$} &  \multicolumn{1}{c}{$\rsTR$} \\
\hline
$\vect{\upmu} / \rsUdt^{-1}$ & $(4.9 \pm 1.2) \times 10^{-3}$ & $(3.1 \pm 0.8) \times 10^{-3}$ & $(5.7 \pm 1.0) \times 10^{-3}$ & $(6.8 \pm 1.2) \times 10^{-3}$ \\
$\vect{\upgamma}$ & $(5.8 \pm 0.7) \times 10^{-1}$ & $(2.2 \pm 0.8) \times 10^{-1}$ & $(7.8 \pm 0.6) \times 10^{-1}$ & \\
$\vect{\upgamma}$ & $(6.0 \pm 0.6) \times 10^{-1}$ & $(2.8 \pm 0.6) \times 10^{-1}$ & & $(7.4 \pm 0.7) \times 10^{-1}$ \\
$\vect{\upbeta} / \rsUdt^{-1}$ & $(7.4 \pm 1.0) \times 10^{-2}$ & $(1.7 \pm 0.4) \times 10^{-2}$ & $(3.9 \pm 0.7) \times 10^{-2}$ & $(2.5 \pm 0.4) \times 10^{-2}$ \\
$\vect{\upxi}$ & $(2.2 \pm 0.6) \times 10^{-1}$ & $(-3.1 \pm 7.4) \times 10^{-2}$ & $(2.5 \pm 0.7) \times 10^{-1}$ & $(9.1 \pm 6.7) \times 10^{-2}$ \\
$\vect{\varsigma}$ & $(3.8 \pm 0.5) \times 10^{-3}$ & $(3.4 \pm 0.6) \times 10^{-3}$ & $(3.7 \pm 0.5) \times 10^{-3}$ & $(5.1 \pm 0.7) \times 10^{-3}$ \\
$\vect{\upeta}$ & $(3.2 \pm 0.9) \times 10^{-2}$ & $(5.2 \pm 0.8) \times 10^{-2}$ & $(3.1 \pm 0.9) \times 10^{-2}$ & $(2.9 \pm 1.0) \times 10^{-2}$ \\
$\vect{\upalpha}$ & $(3.6 \pm 2.0) \times 10^{-1}$ & $2.2 \pm 3.6$ & $(1.6 \pm 2.0) \times 10^{-1}$ & $4.0 \pm 4.1$ \\
\end{tabular}
\end{ruledtabular}
\footnotetext[1]{Left-tail: $x_t < u_{\rsTL} = \rsThreshLeft{}$.}
\footnotetext[2]{Right-tail: $x_t > u_{\rsTR} = \rsThreshRight{}$.}
\end{table*}
\begin{table*}
\caption{
L-BFGS-B parameter estimates ($\pm$ standard errors) for the asymmetric and symmetric common intensity 2T-POT Hawkes models trained on the extreme losses and gains of the S\&P 500 daily log-returns from \rsDateTrainStart{} to \rsDateTrainEnd{}.}
\label{tab:params_mark_ci}
\begin{ruledtabular}
\begin{tabular}{clll}
\multicolumn{1}{c}{}\Tstrut & \multicolumn{2}{c}{$\hat{\vect{\uptheta}}\rsMdlAci{}$} & \multicolumn{1}{c}{$\hat{\vect{\uptheta}}\rsMdlSci{}$}\\
\cmidrule{2-3}\cmidrule{4-4}
\multicolumn{1}{c}{Parameter} & \multicolumn{1}{c}{$\rsTL$} & \multicolumn{1}{c}{$\rsTR$} & \multicolumn{1}{c}{$\rsTA$\footnotemark[3]} \\
\hline
$\mu_{\rsTA} / \rsUdt^{-1}$ & $(7.7 \pm 1.4) \times 10^{-3}$ & & $(8.5 \pm 1.4) \times 10^{-3}$ \\
$\vect{\upgamma}_{\rsTA} $ & $1.2 \pm 0.1$ & $(5.4 \pm 1.0) \times 10^{-1}$ & $(8.3 \pm 0.5) \times 10^{-1}$ \\
$\vect{\upbeta} / \rsUdt^{-1}$ & $(7.6 \pm 1.0) \times 10^{-2}$ & $(1.6 \pm 0.4) \times 10^{-2}$ & $(4.9 \pm 0.5) \times 10^{-2}$ \\
$\vect{\upxi}$ & $(2.2 \pm 0.6) \times 10^{-1}$ & $(-3.2 \pm 6.1) \times 10^{-2}$ & $(1.6 \pm 0.4) \times 10^{-1}$ \\
$\vect{\varsigma}$ & $(3.7 \pm 0.5) \times 10^{-3}$ & $(3.4 \pm 0.6) \times 10^{-3}$ & $(3.5 \pm 0.4) \times 10^{-3}$ \\
$\vect{\upeta}$ & $(3.2 \pm 0.9) \times 10^{-2}$ & $(5.3 \pm 0.8) \times 10^{-2}$ & $(2.2 \pm 0.3) \times 10^{-2}$ \\
$\vect{\upalpha}$ & $(3.6 \pm 1.9) \times 10^{-1}$ & $1.5 \pm 2.4$ & $(7.0 \pm 3.0) \times 10^{-1}$ \\
\end{tabular}
\end{ruledtabular}
\footnotetext[3]{Common-tail: $\left|x_t - \left(u_{\rsTR} + u_{\rsTL}\right)/2\right| > u_{\rsTA} = \left(u_{\rsTR} - u_{\rsTL}\right)/2 = \rsThreshBoth{}$.}
\end{table*}

\subsection{\label{ssec:analysis_likelihood}Likelihood-based inference}

The general bivariate and common intensity 2T-POT Hawkes models ($\vect{\uptheta}\rsMdlBi{}$ and $\vect{\uptheta}\rsMdlAci{}$) are novel descriptions of asymmetric cross-excitement between asymmetric left- and right-tail extreme events; their constrained forms ($\vect{\uptheta}\rsMdlUni{}$ and $\vect{\uptheta}\rsMdlSci{}$) are equivalent to single-tail models that have been applied to financial returns in previous literature \cite{Grothe2014, Gresnigt2015}. We use likelihood-based inference to measure and compare the goodness of fit of each model to the data sample $\mathcal{I}_t$, and so determine which best describes the underlying data generating process of $\mathcal{I}_t$.

The goodness of fit of the model $\vect{\uptheta}$ to $\mathcal{I}_t$ is measured by the log-likelihood function $\ell{(\vect{\uptheta} | \mathcal{I}_t)}$ [defined as \cref{eqn:logL} in \cref{sec:ML}], with higher values of $\ell$ indicating a better fit. The log-likelihood is often quoted as the deviance $-2\ell$, for which lower values are better optimi\ENz{}ed. While the deviance can itself be used to compare the fitness, it is more common to use the Akaike information criterion
\begin{equation}
\label{eqn:AIC}
\mathrm{AIC}{\left(\vect{\uptheta} \middle| \mathcal{I}_t\right)} = 2\mathrm{dim}{\left(\vect{\uptheta}\right)} - 2\ell{(\vect{\uptheta} | \mathcal{I}_t)} ,
\end{equation}
which approximates the expected deviance of a hypothetical new sample that is independent of $\mathcal{I}_t$, and, in doing so, penali\ENz{}es redundant complexity \cite{Wit2012}. An alternative penali\ENz{}ed deviance is the Bayesian information criterion \cite{Wit2012}
\begin{equation}
\label{eqn:BIC}
\mathrm{BIC}{\left(\vect{\uptheta} \middle| \mathcal{I}_t\right)} = \ln{\left[\mathrm{dim}{\left(\mathcal{I}_t\right)}\right]} - 2\ell{(\vect{\uptheta} | \mathcal{I}_t)} .
\end{equation}

\cref{tab:likelihood} lists the penali\ENz{}ed deviance scores for all models in both the training and forecasting periods. A clear hierarchy of fitness emerges from these scores. The decoupled model with no cross-excitation $\vect{\uptheta}\rsMdlUni{}$ yields the worst fit, followed by the symmetric common intensity model $\vect{\uptheta}\rsMdlSci{}$. The novel 2T-POT models with asymmetric interactions provide the best fit, with comparatively little difference between the two: $\vect{\uptheta}\rsMdlAci{}$ is preferred to $\vect{\uptheta}\rsMdlBi{}$ by $\mathrm{AIC}$ and $\mathrm{BIC}$ in the training period, but the opposite is true in the forecasting period.

\begin{table*}
\caption{\label{tab:likelihood}
Deviance and penali\ENz{}ed deviance scores of all 2T-POT Hawkes models against the extreme losses and gains of the S\&P 500 daily log-returns in both the training and forecasting periods.}
\begin{ruledtabular}
\begin{tabular}{ccccccccccc}
\multicolumn{1}{c}{} & \multicolumn{4}{c}{Train (\rsDateTrainStart{} -- \rsDateTrainEnd{})} & \multicolumn{4}{c}{Forecast (\rsDateTrainEnd{} -- \rsDateDataEnd{})}\\
\cmidrule{2-5}\cmidrule{6-9}
\multicolumn{1}{c}{Deviance score} & \multicolumn{1}{c}{$\hat{\vect{\uptheta}}\rsMdlBi{}$} &  \multicolumn{1}{c}{$\hat{\vect{\uptheta}}\rsMdlUni{}$} & \multicolumn{1}{c}{$\hat{\vect{\uptheta}}\rsMdlAci{}$} &  \multicolumn{1}{c}{$\hat{\vect{\uptheta}}\rsMdlSci{}$} & \multicolumn{1}{c}{$\hat{\vect{\uptheta}}\rsMdlBi{}$} &  \multicolumn{1}{c}{$\hat{\vect{\uptheta}}\rsMdlUni{}$} & \multicolumn{1}{c}{$\hat{\vect{\uptheta}}\rsMdlAci{}$} &  \multicolumn{1}{c}{$\hat{\vect{\uptheta}}\rsMdlSci{}$}\\
\hline
$-2\ell{(\hat{\vect{\uptheta}})}$ & 46.42 & 250.30 & 48.43 & 138.85 & -6.72 & 113.06 & -4.44 & 27.43 \\
$\mathrm{AIC}{(\hat{\vect{\uptheta}})}$ & 78.42 & 278.30 & 74.43 & 152.85 & -6.72 & 113.06 & -4.44 & 27.43 \\
$\mathrm{BIC}{(\hat{\vect{\uptheta}})}$ & 160.29 & 349.93 & 140.94 & 188.66 & -0.37 & 118.62 & 0.72 & 30.20 \\
\end{tabular}
\end{ruledtabular}
\end{table*}

The relative fitness of pairs of models is compared directly through the likelihood ratio test \cite{Ruppert2015a}. Specifically, for a given pair of models, $\vect{\uptheta}_0$ and $\vect{\uptheta}_1$, where $\mathrm{dim}{\left(\vect{\uptheta}_0\right)} < \mathrm{dim}{\left(\vect{\uptheta}_1\right)}$, the null hypothesis, $H_0: \ell{(\vect{\uptheta}_0)} = \ell{(\vect{\uptheta}_1)}$, is tested against the alternative, $H_1: \ell{(\vect{\uptheta}_0)} < \ell{(\vect{\uptheta}_1)}$. $H_0$ is rejected when the higher-dimensional (i.e.\ENLA{} more complex) model yields a significantly better fit to the data sample $\mathcal{I}_t$. Since the contributions to the log-likelihood $\ell$ from left- and right-tail events are independent, the likelihood ratio test can be used to compare the relative fitness to each process -- $N_{\rsTL}$, $N_{\rsTR}$, and $N_{\rsTA}$ -- separately.

\cref{tab:LR_test} lists the $p$-values for the likelihood ratio test applied to model pairs with respect to all three processes. The results for $N_{\rsTA}$ -- which correspond to the scores quoted in \cref{tab:likelihood} -- confirm that the classical POT Hawkes models are rejected in fav\ENou{}r of the 2T-POT models with asymmetric interactions at the 95\% significance level.  The results for $N_{\rsTL}$ and $N_{\rsTR}$ show that this is primarily because the latter provide a significantly better fit to the right-tail exceedances events -- supporting the finding that the excitement of $\lambda_{\rsTR}$ is mostly influenced by the history of left-tail events. Notably, $\vect{\uptheta}\rsMdlAci{}$ is never rejected in fav\ENou{}r of $\vect{\uptheta}\rsMdlBi{}$. By directly examining the parameter estimates in \cref{tab:params_mark_bi,tab:params_mark_ci}, we find that the estimated parameters for $\vect{\uptheta}\rsMdlBi{}$ are effectively equivalent to those for $\vect{\uptheta}\rsMdlAci{}$ (i.e.\ENLA{} $\gamma_{\rsTL\rsTL} \approx \gamma_{\rsTR\rsTL} \approx \gamma_{\rsTA\rsTL}/2$ and $\gamma_{\rsTL\rsTR} \approx \gamma_{\rsTR\rsTR} \approx \gamma_{\rsTA\rsTR}/2$), hence, the comparable goodness of fit between the two models. This can also be seen in \cref{fig:SP_process_bi-mark}, where, by visual inspection, $\lambda_{\rsTL} \approx \lambda_{\rsTR}$.  We therefore infer that the arrival of extreme losses and gains is governed by a common conditional intensity, and that this intensity is best approximated by $\lambda_{\rsTA}{( t | \vect{\uptheta}\rsMdlAci{}; \mathcal{I}_t)}$. 

\begin{table}
\caption{\label{tab:LR_test}
Likelihood ratio test $p$-values for the 2T-POT Hawkes models during the training and forecasting periods. $H_0: \ell{\left(\vect{\uptheta}_0\right)} = \ell{\left(\vect{\uptheta}_1\right)}$. $H_1: \ell{\left(\vect{\uptheta}_0\right)} < \ell{\left(\vect{\uptheta}_1\right)}$. Rejections of $H_0$ at the 95\% confidence level are highlighted in bold.}
\begin{ruledtabular}
\begin{tabular}{cccll}
 &  &  & \multicolumn{2}{c}{$p_{\mathrm{LR}}$} \\
\cmidrule{4-5}
Process & $\vect{\uptheta}_0$ & $\vect{\uptheta}_1$ & \multicolumn{1}{c}{Train} & \multicolumn{1}{c}{Forecast} \\
\hline
$N_\rsTL{}$ & $\hat{\vect{\uptheta}}\rsMdlSci{}$ & $\hat{\vect{\uptheta}}\rsMdlAci{}$ & $\pmb{7.5 \times 10^{-8}}$ & $5.2 \times 10^{-2}$ \\
 & $\hat{\vect{\uptheta}}\rsMdlUni{}$ & $\hat{\vect{\uptheta}}\rsMdlAci{}$ & $9.3 \times 10^{-1}$ & $7.5 \times 10^{-2}$ \\
 & $\hat{\vect{\uptheta}}\rsMdlAci{}$ & $\hat{\vect{\uptheta}}\rsMdlBi{}$ & $9.5 \times 10^{-1}$ & $3.5 \times 10^{-1}$ \\
\hline
$N_\rsTR{}$ & $\hat{\vect{\uptheta}}\rsMdlSci{}$ & $\hat{\vect{\uptheta}}\rsMdlAci{}$ & $\pmb{2.4 \times 10^{-8}}$ & $\pmb{3.6 \times 10^{-3}}$ \\
 & $\hat{\vect{\uptheta}}\rsMdlUni{}$ & $\hat{\vect{\uptheta}}\rsMdlAci{}$ & $\pmb{1.9 \times 10^{-40}}$ & $\pmb{1.4 \times 10^{-20}}$ \\
 & $\hat{\vect{\uptheta}}\rsMdlAci{}$ & $\hat{\vect{\uptheta}}\rsMdlBi{}$ & $6.5 \times 10^{-1}$ & $1.0$ \\
\hline
$N_\rsTA{}$ & $\hat{\vect{\uptheta}}\rsMdlSci{}$ & $\hat{\vect{\uptheta}}\rsMdlAci{}$ & $\pmb{2.5 \times 10^{-17}}$ & $\pmb{1.7 \times 10^{-5}}$ \\
 & $\hat{\vect{\uptheta}}\rsMdlUni{}$ & $\hat{\vect{\uptheta}}\rsMdlAci{}$ & $\pmb{8.1 \times 10^{-46}}$ & $\pmb{2.2 \times 10^{-27}}$ \\
 & $\hat{\vect{\uptheta}}\rsMdlAci{}$ & $\hat{\vect{\uptheta}}\rsMdlBi{}$ & $5.7 \times 10^{-1}$ & $5.2 \times 10^{-1}$ \\
\end{tabular}
\end{ruledtabular}
\end{table}

For developed market indices, such as the S\&P 500, it is known that extreme returns on consecutive trading days exhibit an asymmetric sign persistence: extreme gains are persistent, meaning they are more likely to be directly followed by another extreme gain than by an extreme loss; conversely, extreme losses are reversive, meaning they are also more likely followed by an extreme gain \cite{Chicheportiche2014}. However, these persistences only exist over a single trading day -- a timescale that is too short to be detected by a Hawkes model with inverse decay constants $\beta^{-1} > 10^{1} \: \rsUdt$. It is therefore intuitive that extreme losses and gains are found to share a common arrival intensity.

Having concluded that the common intensity model best describes the data $\mathcal{I}_t$, we examine the values of its estimated parameters $\hat{\vect{\uptheta}}\rsMdlAci{}$ when fitted to these data, as listed in \cref{tab:params_mark_ci}. We observe significant asymmetries in the values estimated for the two tails. First, there is an asymmetry in the excitation vector $\vect{\upgamma}_{\rsTA}$, such that $\gamma_{\rsTA\rsTL}/\gamma_{\rsTA\rsTR} = 2.2 \pm 0.5$. This means that, on average, extreme losses trigger more than twice as many daughter events (from either tail) as extreme gains. At the same time, the ratio between the decay constants, $\beta_{\rsTL}/\beta_{\rsTR} = 4.6 \pm 1.2$, means that the excitation from losses decays significantly faster, and, therefore, that this excitement is more concentrated in time to the immediate aftermath of the mother event's arrival. These asymmetries are consistent with previous studies of developed market indices, which have found that extreme daily losses are a better predictor of future daily extremes (gain or loss) than are extreme daily gains \cite{Embrechts2011}. \cref{sec:CV} examines in detail whether these asymmetries can be explained by invoking volatility clustering and the leverage effect as described by GARCH models. We find that the leverage effect as described by the GJR-GARCH model \cite{Glosten1993} can account for the asymmetry in $\vect{\upgamma}_{\rsTA}$ but not the asymmetry in $\vect{\upbeta}$. The latter, a novel insight of the 2T-POT Hawkes model, suggests a more complex data generating process for log-returns, in which the leverage effect is more pronounced at shorter timescales.

\subsection{\label{ssec:analysis_residual}Residual analysis}

We further assess the performance of the 2T-POT Hawkes models at describing the arrival process through the residual analysis technique developed by Ogata \cite{Ogata1988}. 

If the continuous time arrivals of the point process $N_{i}{\left(t\right)}$ are described by the conditional intensity $\lambda_{i}{\left(t\right)}$, then, in the residual time 
\begin{equation}
\label{eqn:tau}
\tau_{i}{\left(t\right)} = \int_0^t{\lambda_{i}{\left(t'\right)} d{t'}} ,
\end{equation}
the residual process $N_{i}{\left(\tau_i\right)}$ is a homogeneous unit Poisson process and the residual interarrivals $\Delta{\tau}_{i,k_{i}} = \tau_{i}{\left(t_{k_i}\right)} - \tau_{i}{\left(t_{{k_i}-1}\right)}$ are therefore i.i.d.\@ unit exponential random variables. If event arrivals instead occur in discrete time with a minimum time-step $\delta{t}$, then these expected distributions are asymptotic in the limit $\lambda_i \delta{t} \to 0$.

The bivariate model natively yields separate residual processes for left- and right-tail exceedances. Generically, it can be shown that $\lambda_{\rsTA} \equiv \left< \vect{1} \middle| \vect{\uplambda}\right> \equiv \lambda_{\rsTL} + \lambda_{\rsTR}$, and so a residual time for exceedances from both tails, $\tau_{\rsTA}$, is trivial to derive from $\vect{\uplambda}$. It is less trivial to derive $\tau_{\rsTL}$ and $\tau_{\rsTR}$ from the common intensity model, since there is no inverse function to calculate $\lambda_{\rsTL}$ and $\lambda_{\rsTR}$ from $\lambda_{\rsTA}$. Instead, the residual interarrivals of these processes are derived from the probability of an event occurring in $N_{\rsTA}$ and then being stochastically drawn from either tail, with relative frequency $\mathbb{E}{[N_{\rsTL}/N_{\rsTR}]} = \exp{\left(-w_{\rsTA}\right)}$:
\begin{align}
\label{eqn:delta_tau_split}
\Delta{\tau_{\rsTO, k_{\rsTO}}} &= \Delta{\tau_{\rsTA,k_{\rsTO}}} \mp\frac{w_{\rsTA}}{2} 
\nonumber\\
&+ \ln{\left[\pm\sinh{\frac{w_{\rsTA}}{2}} + \sqrt{\sinh^2{\frac{w_{\rsTA}}{2}} + e^{-\Delta{\tau_{\rsTA,k_{\rsTO}}}}}\right]}.
\end{align}
If $w_{\rsTA} = 0$, \cref{eqn:delta_tau_split} reduces to $\Delta{\tau_{\rsTO, k_{\rsTO}}} = \Delta{\tau_{\rsTA,k_{\rsTO}}}/2$.

\begin{figure*}
\includegraphics{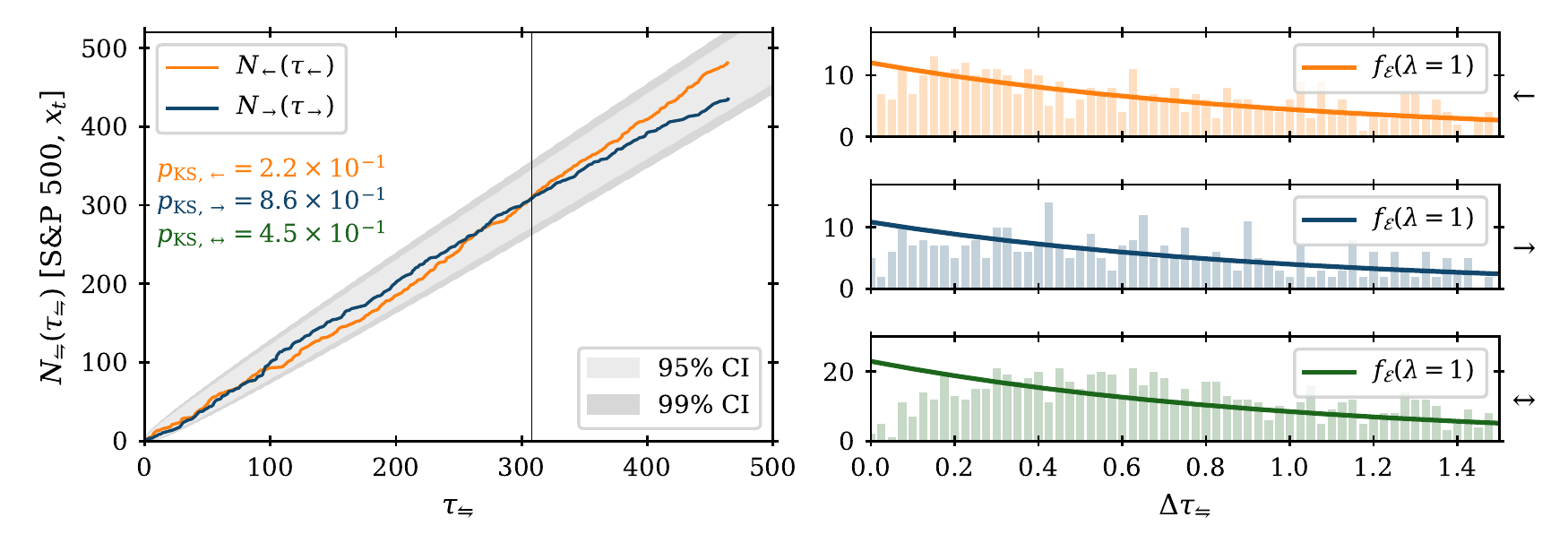}
\caption{
\label{fig:SP_N_delta_tau_ci-mark_KS}
Residual arrival processes of extreme S\&P 500 daily log-returns under $\vect{\uptheta}\rsMdlAci{}$. Left panel: left- (light orange) and right-tail (dark blue) exceedance count against residual time. The KS test is used against the null hypothesis, $H_0:\lambda_{\rsTO}{\left(t\right)} = \lambda_{\rsTO}{\left(t \middle| \vect{\uptheta}\rsMdlAci{}; \mathcal{I}_t\right)}$: the gr\ENe{}y shaded areas show the 95\% (lighter) and 99\% (darker) KS confidence intervals; also shown are the training period KS $p$-values for $N_{\rsTL}$, $N_{\rsTR}$ and $N_{\rsTA}$, none of which are rejected at the 95\% confidence level. The vertical black line marks the end of the training period on \rsDateTrainEnd{}. Right panels: histogram of the residual interarrivals for exceedances from the left tail (top-right panel, light orange), right tail (middle-right panel, dark blue), and both tails (bottom-right panel, green); the solid lines show the unit exponential distribution expected under $H_0$.
} 
\end{figure*}

The residual processes under the common intensity model $\vect{\uptheta}\rsMdlAci{}$ are shown in \cref{fig:SP_N_delta_tau_ci-mark_KS}; this is also representative of the corresponding residual processes under $\vect{\uptheta}\rsMdlBi{}$, due to the approximate equivalence of the estimated parameters, as discussed in \cref{ssec:analysis_likelihood}. 
In contrast to \cref{fig:SP_N_delta_t_KS}, we observe that the residual processes are approximately unit Poisson, and, therefore, that the true conditional intensities are well approximated by $\vect{\uplambda}{\left(t \middle| \vect{\uptheta}\rsMdlAci{}; \mathcal{I}_t\right)}$. A minor, but notable exception is seen in the bottom-right panel of  \cref{fig:SP_N_delta_tau_ci-mark_KS}, where there is a decline in the observed frequency relative to expectation in the limit $\Delta{\tau_{\rsTA}} \to 0$, i.e.\ENLA{} there are fewer than expected instances of exceedance events arriving almost simultaneously in $N_{\rsTA}{(\tau_{\rsTA})}$. This is a discreti\ENz{}ation error that arises because the arrivals occur in discrete time: since $\mathbb{E}{[\lambda_{\rsTA}/\lambda_{\rsTL}]} = \mathbb{E}{[\lambda_{\rsTA}/\lambda_{\rsTR}]} = 2$, $\lambda_{\rsTA}$ is always further from the asymptotic limit in which the approximation of continuous time arrivals is valid; this suppresses severe under-forecasting, because the interarrival times cannot be less than discrete time step $\delta{t}$.

We note that there are persistent inequalities in the arrival frequencies of losses versus gains in the residual time. This is especially pronounced in the forecasting period, i.e.\ENLA{} for data outside of the sample whose symmetric quantiles were used to define the thresholds. While our method of threshold definition guarantees an equal number of left- and right-tail exceedances within the training period, we find a 168 to 124 split within the forecasting period. This reflects the use of constant thresholds without prior normali\ENz{}ation of the original time series: while this approach avoids interference with the short timescale signals of self-excitement, it cannot account for long timescale changes in the distribution of returns. Future work may look to incorporate these long-term trends while minimi\ENz{}ing such interference.

We test the null hypothesis that the true conditional intensity and the conditional intensity approximated by the Hawkes model are the same, i.e.\ENLA{} $H_0:\lambda_{i}{\left(t\right)} = \lambda_{i}{\left(t \middle| \vect{\uptheta}; \mathcal{I}_t\right)}$, by performing the Kolmogorov-Smirnov (KS) test \cite{Lopes2011} on the null hypothesis that the residual process derived from $\lambda_{i}{\left(t \middle| \vect{\uptheta}; \mathcal{I}_t\right)}$ is unit Poisson. \cref{tab:KS_mark} shows the $p$-values of this test performed for each tail intensity -- $\lambda_{\rsTL}$, $\lambda_{\rsTR}$, and $\lambda_{\rsTA}$ -- within both the training and forecasting periods. 
\begin{table}
\caption{\label{tab:KS_mark}
Kolmogorov-Smirnov (KS) test $p$-values for the residual processes of exceedance arrivals from either and both tails under the 2T-POT Hawkes models during the training and forecasting periods. $H_{0}: \lambda_i{\left(t\right)} = \lambda_i{\left(t \middle| \vect{\uptheta}\right)}$. $H_{1}: \lambda_i{\left(t\right)} \neq \lambda_i{\left(t \middle| \vect{\uptheta}\right)}$. Rejections of $H_0$ at the 95\% confidence level are highlighted in bold.}
\begin{ruledtabular}
\begin{tabular}{cllllll}
 & \multicolumn{6}{c}{$p_{\mathrm{KS}}$} \\
 \cmidrule{2-7}
 & \multicolumn{3}{c}{Train} &  \multicolumn{3}{c}{Train and Forecast} \\
\cmidrule{2-4}\cmidrule{5-7}
 Model & $\lambda_{\rsTL}$ & $\lambda_{\rsTR}$ & $\lambda_{\rsTA}$ & $\lambda_{\rsTL}$ & $\lambda_{\rsTR}$ & $\lambda_{\rsTA}$ \\
\hline
$\vect{\uptheta}{\rsMdlBi{}}$& $0.113$ & $0.946$ & $0.416$ & $0.206$ & $\pmb{0.014}$ & $0.671$ \\
$\vect{\uptheta}{\rsMdlUni{}}$& $0.194$ & $0.192$ & $0.077$ & $0.444$ & $0.379$ & $0.223$ \\
$\vect{\uptheta}{\rsMdlAci{}}$& $0.217$ & $0.857$ & $0.449$ & $0.287$ & $\pmb{0.029}$ & $0.702$ \\
$\vect{\uptheta}{\rsMdlSci{}}$& $0.098$ & $0.867$ & $0.205$ & $0.061$ & $0.168$ & $0.428$ \\
\end{tabular}
\end{ruledtabular}
\end{table}
There are few rejections at the 95\% significance level and there are no such rejections for tests performed on $\lambda_{\rsTA}$. We note that, with one exception, the KS $p$-values under $\vect{\uptheta}\rsMdlAci{}$ are higher than under $\vect{\uptheta}\rsMdlBi{}$, supporting the conclusion in \cref{ssec:analysis_likelihood} that the common intensity model is the optimal choice for the S\&P 500 data set.

\begin{figure}[t]
\includegraphics{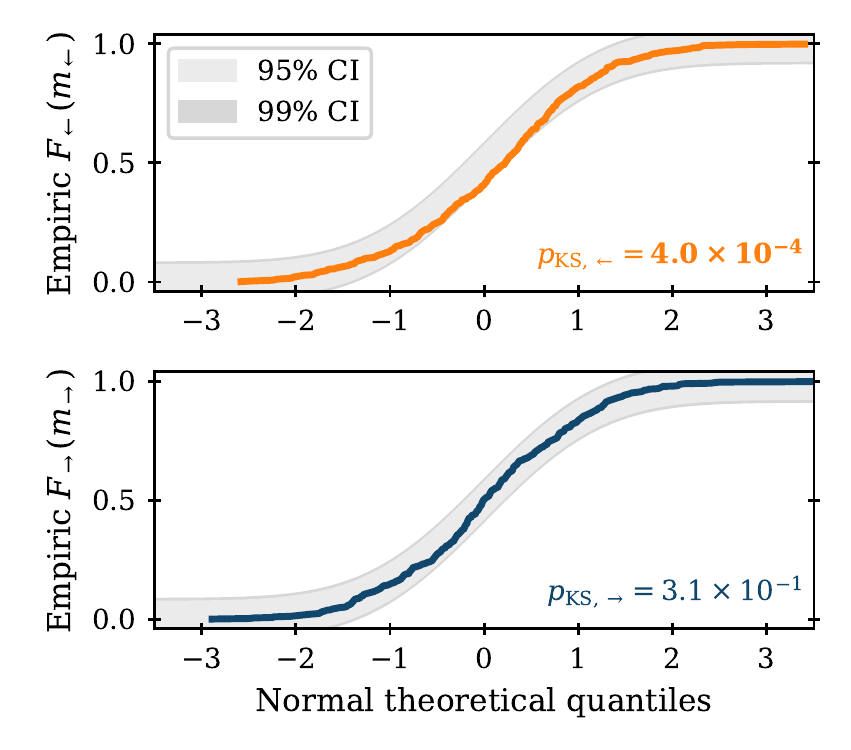}
\caption{
\label{fig:SP_m_KS_ci-mark}
Kolmogorov-Smirnov (KS) test for left- (top panel, light orange) and right-tail (bottom panel, dark blue) residual excess magnitudes for the S\&P 500 extreme daily log-returns under $\vect{\uptheta}\rsMdlAci{}$ compared against the unit exponential distribution. The gr\ENe{}y shaded areas show the 95\% (lighter) and 99\% (darker) KS confidence intervals.
}
\end{figure}

We complement the residual analysis of the arrivals processes with a residual analysis of the excess magnitudes. If the excesses $\left\{m_{k_\rsTO}\right\}$ are distributed according to the conditional GPD specified in \cref{eqn:F_GPD}, then the residual excess magnitudes,
\begin{align}
\label{eqn:expon_gpd_transform}
\mathcal{E}{\left(m_{k_\rsTO}\right)} = 
\begin{cases}
	\xi_\rsTO^{-1} \ln{\left[1 + \xi_\rsTO \frac{m_{k_\rsTO}}{\sigma_{\rsTO}{\left(t_{k_\rsTO}\right)}}\right]}, & \xi_\rsTO \neq 0,  \\
	m_{k_\rsTO}/\sigma_{\rsTO}{\left(t_{k_\rsTO}\right)}, & \xi_{\rsTO} = 0,  \\
\end{cases}
\end{align}
are approximately i.i.d.\@ unit exponential random variables. In \cref{fig:SP_m_KS_ci-mark}, we compare in the top and bottom panel, respectively, the left- and right-tail residual excess magnitudes for our extreme events under the model $\vect{\uptheta}\rsMdlAci{}$ to the unit exponential distribution. We observe a very good agreement between the two apart from in the vicinity of the median where the left-tail residuals cross the KS 99\% confidence interval bounds. We attribute this to an underestimation of the left-tail conditional intensity $\lambda_{\rsTL}$ (and, therefore, of the left-tail conditional scale parameter $\sigma_{\rsTL}$) at the onset of high-activity clusters within the forecasting period. Notably, these clusters correspond to the Global Financial Crisis (2007-9) and the Covid-19 pandemic (2020). This result is consistent with there being additional sources of non\ENhyphNon{}constant exogenous intensity that we have not accounted for here.

\begin{figure}[t]
\includegraphics{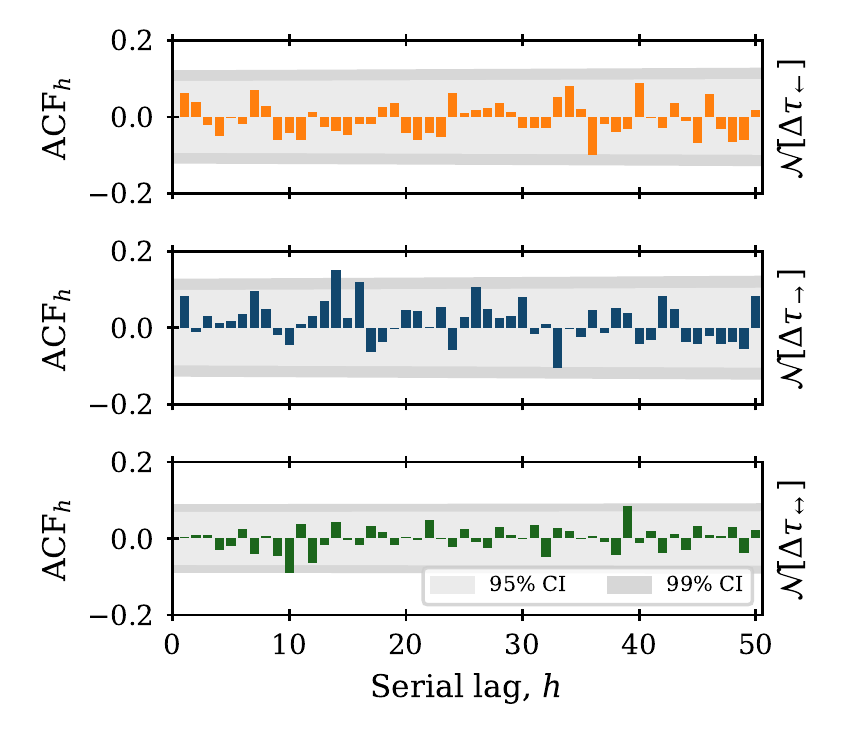}
\caption{
\label{fig:SP_corr_ci-mark_c}
Correlograms for the transformed residual interarrivals of S\&P 500 extreme daily log-returns under $\vect{\uptheta}\rsMdlAci{}$; derived from $N_{\rsTL}$ (top panel, light orange), $N_{\rsTR}$ (middle panel, dark blue), and $N_{\rsTA}$ (bottom panel, green). The gr\ENe{}y shaded areas show the 95\% (lighter) and 99\% (darker) confidence intervals for $\mathrm{ACF}_h = 0$, respectively.
}
\end{figure}
\begin{figure*}
\includegraphics{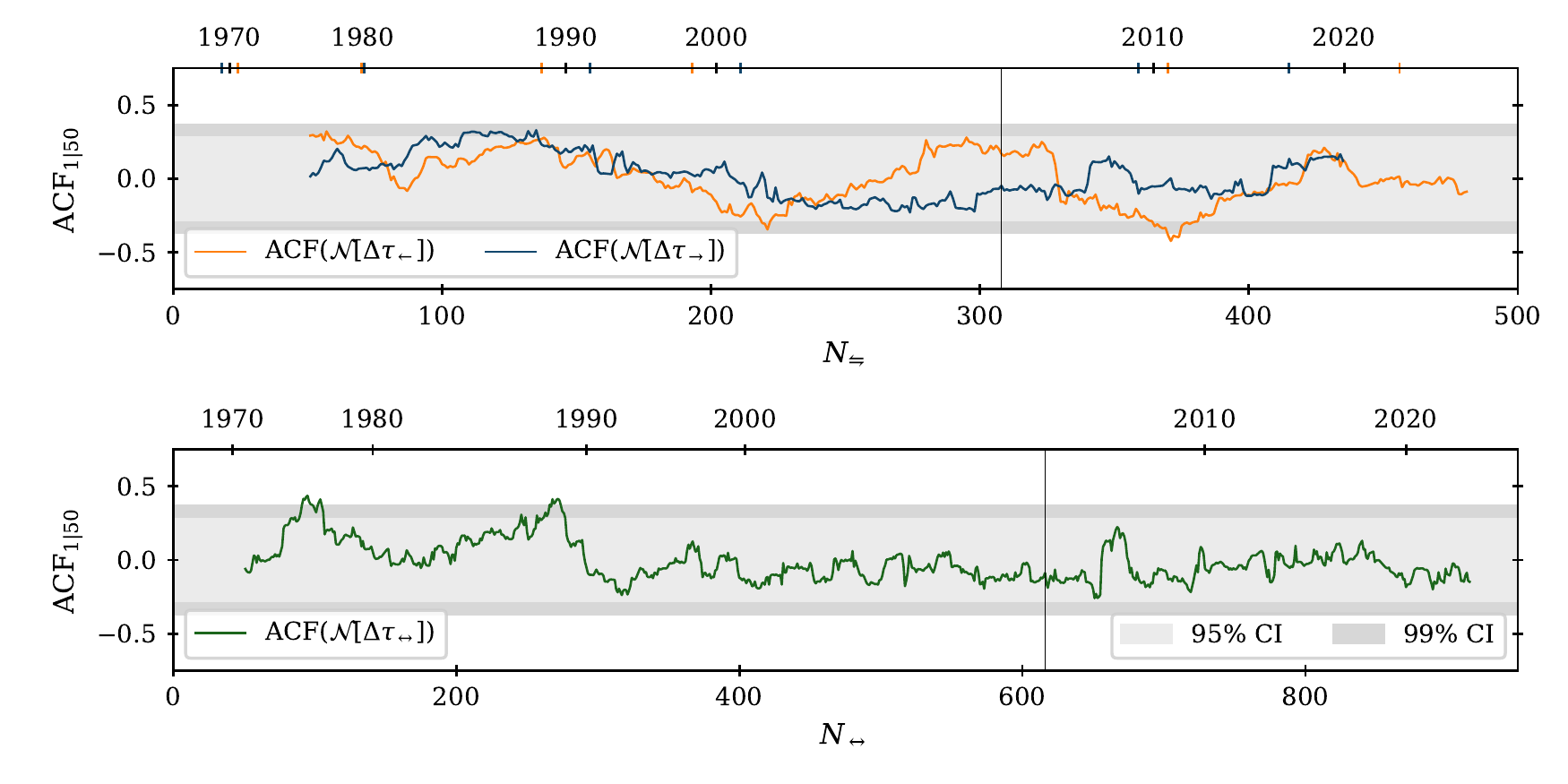}
\caption{
\label{fig:SP_ACF_ci-mark}
Rolling-window (length 50) lag-1 autocorrelation for transformed residual interarrivals of extreme S\&P 500 daily log returns under $\vect{\uptheta}\rsMdlAci{}$; derived from $N_{\rsTL}$ (top panel, light orange), $N_{\rsTR}$ (top panel, dark blue) and $N_{\rsTA}$ (bottom panel, green). The gr\ENe{}y shaded areas show the 95\% (lighter) and 99\% (darker) confidence intervals for $\mathrm{ACF}_{1|50} = 0$, respectively; the vertical black line marks the end of the training period on \rsDateTrainEnd{}. 
}
\end{figure*} 

We investigate serial dependence within the residual interarrivals as a signal of systemic under- or over-forecasting. This has implications for the forecasting ability of the model in practice, but it could also herald additional dynamics within the true underlying arrival process that are not captured by the model. For this analysis, we transform the residual interarrivals from an expected unit exponential distribution to an expected unit normal distribution via the operation
\begin{align}
\mathcal{N}{\left(\Delta{\tau_{i,k_i}}\right)} 
&= F_{\mathrm{normal}}^{-1}{F_{\mathrm{expon}}}{\left(\Delta{\tau_{i,k_i}}\right)} 
\nonumber\\
&= \sqrt{2} \: \mathrm{erf}^{-1}{\left(1 - 2\exp{\left[-\Delta{\tau_{i,k_i}}\right]}\right)} ,
\label{eqn:norm_expon_transform}
\end{align}
where $\mathrm{erf}^{-1}$ is the inverse error function. Thus, if the exceedance event at time $t_{k_{i}}$ is under-forecast (i.e.\ENLA{} arrived sooner than expected) by the model, then $\mathcal{N}{\left(\Delta{\tau_{i,k_i}}\right)} < 0$.

\cref{fig:SP_corr_ci-mark_c} shows that, over the full data sample, there is negligible autocorrelation within the normali\ENz{}ed residual interarrivals under $\vect{\uptheta}\rsMdlAci{}$ (and, therefore, under $\vect{\uptheta}\rsMdlBi{}$). Conversely, \cref{fig:SP_ACF_ci-mark} shows peaks of statistically significant locali\ENz{}ed autocorrelation. For the residual interarrivals of $N_{\rsTA}$, three notable peaks of positive lag-1 autocorrelation are observed: these follow the high-activity clusters corresponding to the 1973-4 stock market crash, Black Monday (1987), and the Global Financial Crisis (2007-9). We infer this to be a signal of systemic overestimation of the conditional intensity (i.e.\ENLA{} systemic over-forecasting) in the latter stages of high activity regimes. We speculate that this is also a consequence of neglecting significant additional sources of non\ENhyphNon{}constant exogenous intensity at the start of these regimes: without these sources, all excess intensity must be attributed to endogenous self-excitement alone; this leads to an overestimation of the branching matrix elements, which then works against the relaxation of the conditional intensity at the end of high activity clusters. Rather than being the mechanism by which the system reaches the excited state, self-excitement may more so be the mechanism by which the excited state persists, having been initially instigated by a sudden increase in exogenous intensity corresponding to either impactful news or other complex dynamics within the market.

\section{\label{sec:summary}Summary}

We have developed a two-tailed peaks-over-threshold Hawkes model that captures asymmetric self- and cross-excitation between the left- and right-tail extremes based upon a common conditional intensity. Such a model provides a way to measure and describe self-exciting processes with more than one mutually exclusive but interacting types of extreme behavi\ENou{}rs. When compared to its symmetric version as well as a bivariate model in which each tail contributes to either coupled or decoupled distinct point processes, our model, applied to daily log-returns of the S\&P 500 index, was found to provide the most parsimonious fit to the data as measured by penali\ENz{}ed deviance.

By accounting for asymmetric interactions between the tails, our model finds that, for the S\&P 500 daily log-returns, extreme losses trigger on average more than twice as many daughter events as do extreme gains. The excitation from losses is also found to decay more than four times as quickly as that from gains. While the former of these asymmetries can be explained by invoking the GJR-GARCH leverage effect, the latter is a novel insight of the 2T-POT Hawkes model that provides evidence for the leverage effect being more pronounced at shorter timescales. The greater, more immediate impact of losses is consistent with the greater psychological weight assigned to them by human agents, i.e.\ENLA{} this result reflects a negativity bias wherein negative events generally provoke a stronger response than equivalent positive events \cite{Baumeister2001, Rozin2001}. 

Beyond the demonstrated application to financial data, we anticipate possible extensions of our model to other drift-diffusion-like processes in which a clustering of extreme fluctuations is observed.

\begin{acknowledgements}
We thank the anonymous referees for feedback and suggestions that helped us to enhance the clarity and contextuali\ENz{}ation of our findings. M.F.T. acknowledges support from EPSRC (UK) Grant No. EP/R513155/1 and CheckRisk LLP. 
\end{acknowledgements}

\appendix

\section{\label{sec:CV}Volatility clustering and leverage effect}

When exceedance events are defined against fixed thresholds, a clustering of exceedance arrivals can arise as a consequence of non\ENhyphNon{}constant variance in $x_t$. Volatility clustering is a styli\ENz{}ed fact of financial returns: it states that the standard deviation of log-returns (known as the volatility) is non\ENhyphNon{}constant and exhibits significant positive autocorrelation. Another styli\ENz{}ed fact -- the leverage effect -- states that volatility increases when returns become more negative \cite{Cont2001, Ruppert2015a}. This Appendix examines whether the asymmetric self-excitement reported in this paper can be attributed to these well-established features of financial data.

GARCH models have become the standard model for log-returns in financial engineering due to their parsimonious description of these two styli\ENz{}ed facts \cite{Ruppert2015a}. These models take the form of a conditional volatility process where log-returns are generated as
\begin{equation}
\label{eqn:cond_vol}
x_t = \mu + \sigma_t \epsilon_t ,
\end{equation}
where $\mu$ is the unconditional mean, $\sigma_t$ is the conditional volatility, and $\epsilon_t$ is random noise drawn from a parametric distribution with zero mean and unit variance. In the $\mathrm{GARCH}{(p,o,q)}$ model, the conditional variance $\sigma_t^2$ is described by an autoregressive moving average (ARMA) process
\begin{align}
\sigma_t^2 = \omega &+ \sum_{i=1}^{p}{\alpha_{i}{\left(\sigma_{t-i}\epsilon_{t-i}\right)}^2} + \sum_{j=1}^{q}{\beta_{j}\sigma_{t-j}^2} \nonumber \\
&+ \sum_{k=1}^{o}{\gamma_{k}{\left(\sigma_{t-k}\epsilon_{t-k}\right)}^2 I_{t-k}} ,
\label{eqn:GARCH}
\end{align}
where $\omega$ is the minimum conditional variance and $\{\alpha_i, \beta_j, \gamma_k\}$ are the ARCH coefficients. When $o \neq 0$, \cref{eqn:GARCH} is a GJR-GARCH model \cite{Glosten1993} that accounts for the leverage effect through the indicator function
\begin{equation}
\label{eqn:GARCH_indicator}
I_t = 
\begin{cases}
	1, & \epsilon_{t} < 0,  \\
	0, & \epsilon_{t} \geq 0, \\
\end{cases}
\end{equation}
If the data $x_t$ is generated by the specified conditional volatility model, then the normali\ENz{}ed residuals $z_t$ are i.i.d.\@ unit white noise, i.e.\ENLA{} 
\begin{equation}
\label{eqn:z_t}
z_t = F_{\mathrm{normal}}^{-1}{F_{\epsilon}}{\left[\left(x_t - \mu\right)/\sigma\right]} \sim \mathcal{N}{(0,1)} .
\end{equation}

In this Appendix, we consider the $\mathrm{GARCH}{(p=1,o,q=1)}$ model with ($o=1$) and without ($o=0$) the leverage effect. Hereafter, we label variants of this model as ${\mathrm{G}{(o,.)}}$, where the second argument is a letter specifying the parametric error distribution: $\mathcal{N}$ denotes the unit normal distribution [$\epsilon_t \sim \mathcal{N}{(0,1)}$] and $t$ denotes the unit Student's t distribution [$\epsilon_t \sim t{(0,1,\nu)}$] \cite{Ruppert2015a}. 

\begin{figure}[t]
\includegraphics{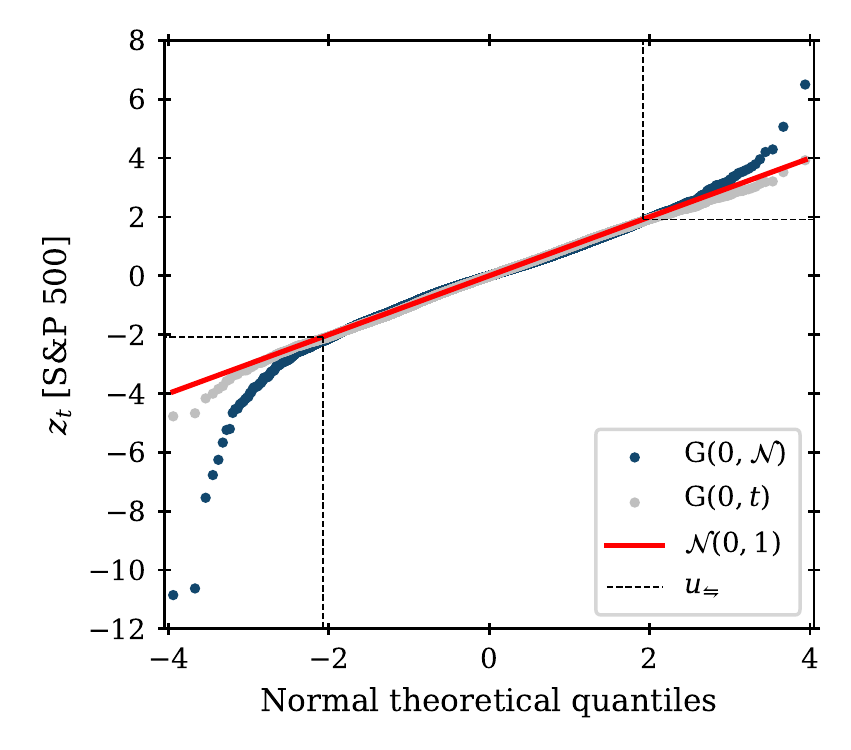}
\caption{
\label{fig:SP_cv_z_G101_qq}
Normal quantile plot of the S\&P 500 GARCH normali\ENz{}ed residuals $z_t$ in the training period (\rsDateTrainStart{} to \rsDateTrainEnd{}) under $\mathrm{G}{(0,\mathcal{N})}$ (dark blue) and $\mathrm{G}{(0,t)}$ (light gr\ENe{}y). The exceedance thresholds (black, dashed) approximately coincide with the visual divergence of the training period residuals from the unit normal distribution $\mathcal{N}{(0,1)}$ (red).  
}
\end{figure}

To examine whether the asymmetric self-excitement of S\&P 500 log-returns reported in this paper can be attributed to volatility clustering and the GJR-GARCH leverage effect alone, we fit the 2T-POT Hawkes model to the normali\ENz{}ed residuals of four GARCH models -- ${\mathrm{G}{(0,\mathcal{N})}}$, ${\mathrm{G}{(0,t)}}$, ${\mathrm{G}{(1,\mathcal{N})}}$, and ${\mathrm{G}{(1,t)}}$ -- fitted to the S\&P 500 data (see \cref{tab:GARCH} for the fitted GARCH parameters). 
\begin{table}
\caption{\label{tab:GARCH}
GARCH model parameter estimates ($\pm$ standard errors) and AIC penali\ENz{}ed deviance for the S\&P 500 daily log-returns from \rsDateTrainStart{} to \rsDateTrainEnd{}.}
\begin{ruledtabular}
\begin{tabular}{cll}
 Parameter & ${\mathrm{G}{(0,\mathcal{N})}}$ & ${\mathrm{G}{(0,t)}}$ \\
\hline
$\mu$ & $(4.5 \pm 0.6) \times 10^{-4}$ & $(4.8 \pm 0.6) \times 10^{-4}$ \\
$\omega$ & $(6.1 \pm 0.9) \times 10^{-5}$ & $(4.8 \pm 0.9) \times 10^{-5}$ \\
$\alpha_1$ & $(8.0 \pm 0.5) \times 10^{-2}$ & $(7.0 \pm 0.5) \times 10^{-2}$ \\
$\beta_1$ & $(9.2 \pm 0.0) \times 10^{-1}$ & $(9.3 \pm 0.1) \times 10^{-1}$ \\
$\nu$ & & $7.5 \pm 0.5$ \\
$\mathrm{AIC}$ & $29402.3$ & $28832.5$ \\
\hline
 Parameter & ${\mathrm{G}{(1,\mathcal{N})}}$ & ${\mathrm{G}{(1,t)}}$ \\
\hline
$\mu$ & $(3.0 \pm 0.6) \times 10^{-4}$ & $(3.7 \pm 0.6) \times 10^{-4}$ \\
$\omega$ & $(7.3 \pm 1.0) \times 10^{-5}$ & $(5.5 \pm 0.9) \times 10^{-5}$ \\
$\alpha_1$ & $(3.1 \pm 0.4) \times 10^{-2}$ & $(2.7 \pm 0.4) \times 10^{-2}$ \\
$\gamma_1$ & $(8.4 \pm 0.7) \times 10^{-2}$ & $(8.2 \pm 0.8) \times 10^{-2}$ \\
$\beta_1$ & $(9.2 \pm 0.0) \times 10^{-1}$ & $(9.3 \pm 0.0) \times 10^{-1}$ \\
$\nu$ & & $8.0 \pm 0.5$ \\
$\mathrm{AIC}$ & $29199.0$ & $28679.6$ \\
\end{tabular}
\end{ruledtabular}
\end{table}
The exceedance thresholds in $z_t$ for the 2T-POT Hawkes model are defined by the \fpeval{(100*\rsQuantileLeft)}\% and \fpeval{(100*\rsQuantileRight)}\% sample quantiles of $z_t$ in the training period. \cref{fig:SP_cv_z_G101_qq} shows that, just like for the log-returns $x_t$ in \cref{fig:SP_x_qq}, this is approximately where the marginal distribution of training period normali\ENz{}ed residuals diverges from the normal distribution.
\begin{figure*}
\includegraphics{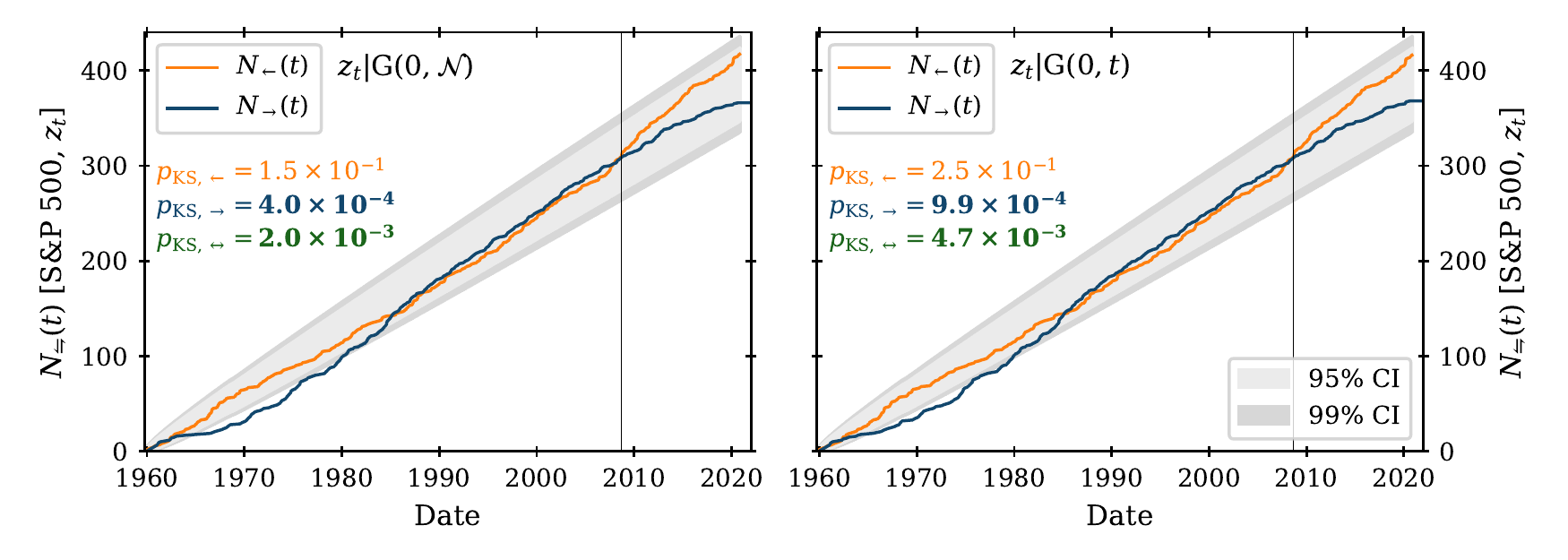}
\caption{
\label{fig:SP_cv_z_tt_G101_N_delta_t}
Arrival processes of extreme left- (light orange) and right-tail (dark blue) S\&P 500 normali\ENz{}ed residuals under $\mathrm{G}{(0,\mathcal{N})}$ (left panel) and $\mathrm{G}{(0,t)}$ (right panel). The KS test is used against the null hypothesis implied by i.i.d.\@ normali\ENz{}ed residuals, i.e.\ENLA{} $H_0: d{N_{\rsTO}}/d{t} \equiv \lambda_{\rsTO}{\left(t\right)} = 0.025 \: \rsUdt^{-1}$: the gr\ENe{}y shaded areas show the 95\% (lighter) and 99\% (darker) KS confidence intervals; also shown are the training period KS $p$-values for $N_{\rsTL}$, $N_{\rsTR}$ and $N_{\rsTA}$, with rejections of the null at the 95\% confidence level highlighted in bold. The vertical black line marks the end of the training period on \rsDateTrainEnd{}. 
}
\end{figure*}
\begin{table*}
\caption{
Fitted parameters ($\pm$ standard errors) for the 2T-POT Hawkes bivariate model $\vect{\uptheta}\rsMdlBi$ with the constraints $\vect{\upeta}=0=\vect{\upalpha}$ fitted to unit white noise $z_t | \mathcal{N}{(0,1)}$ and the normali\ENz{}ed residuals of the S\&P 500 daily log-returns under the GARCH models $z_t | {\mathrm{G}{(0,\mathcal{N})}}$, $z_t | {\mathrm{G}{(0,t)}}$, $z_t | {\mathrm{G}{(1,\mathcal{N})}}$, and $z_t | {\mathrm{G}{(1,t)}}$. Standard errors for $z_t | \mathcal{N}{(0,1)}$ are obtained through \num{10000} simulated series.}
\label{tab:params_cv_z_tt}
\begin{ruledtabular}
\begin{tabular}{clllll}
Parameter & $z_t | \mathcal{N}{(0,1)}$ & $z_t | {\mathrm{G}{(0,\mathcal{N})}}$ & $z_t | {\mathrm{G}{(0,t)}}$ & $z_t | {\mathrm{G}{(1,\mathcal{N})}}$ & $z_t | {\mathrm{G}{(1,t)}}$\\
\hline
$u_{\rsTL}$ & $-1.959$ & $-2.072$ & $-2.033$ & $-2.031$ & $-2.004$ \\
$u_{\rsTR}$ & $1.959$ & $1.915$ & $1.895$ & $1.934$ & $1.909$ \\
$\mu_{\rsTL} / \rsUdt^{-1}$ & $2.5 \times 10^{-2}$ & $(2.1 \pm 0.1) \times 10^{-2}$ & $(2.1 \pm 0.1) \times 10^{-2}$ & $(2.3 \pm 0.9) \times 10^{-2}$ & $(2.5 \pm 0.7) \times 10^{-2}$ \\
$\mu_{\rsTR} / \rsUdt^{-1}$ & $2.5 \times 10^{-2}$ & $(2.2 \pm 0.1) \times 10^{-2}$ & $(2.2 \pm 0.1) \times 10^{-2}$ & $(6.8 \pm 3.0) \times 10^{-3}$ & $(7.1 \pm 3.1) \times 10^{-3}$ \\
$\gamma_{\rsTL\rsTL}$ & $0$ & $(4.7 \pm 2.7) \times 10^{-2}$ & $(4.9 \pm 2.7) \times 10^{-2}$ & $(1.6 \pm 3.0) \times 10^{-1}$ & $(5.4 \pm 29.6) \times 10^{-2}$ \\
$\gamma_{\rsTL\rsTR}$ & $0$ & $0.0 \pm 0.0$ & $0.0 \pm 0.0$ & $(1.8 \pm 2.3) \times 10^{-1}$ & $(2.0 \pm 0.9) \times 10^{-1}$ \\
$\gamma_{\rsTR\rsTL}$ & $0$ & $0.0 \pm 0.0$ & $0.0 \pm 0.0$ & $0.0 \pm 0.0$ & $0.0 \pm 0.0$ \\
$\gamma_{\rsTR\rsTR}$ & $0$ & $0.0 \pm 0.0$ & $0.0 \pm 0.0$ & $(6.9 \pm 1.1) \times 10^{-1}$ & $(6.8 \pm 1.2) \times 10^{-1}$ \\
$\beta_{\rsTL} / \rsUdt^{-1}$ & $0$ & $(2.8 \pm 2.0) \times 10^{-1}$ & $(2.8 \pm 1.8) \times 10^{-1}$ & $(3.0 \pm 3.0) \times 10^{-3}$ & $(3.4 \pm 5.3) \times 10^{-3}$ \\
$\beta_{\rsTR} / \rsUdt^{-1}$ & $0$ & $(2.1 \pm 0.0) \times 10^{-1}$ & $(3.1 \pm 0.0) \times 10^{-2}$ & $(1.6 \pm 0.8) \times 10^{-3}$ & $(1.6 \pm 0.8) \times 10^{-3}$ \\
$\xi_{\rsTL}$ & $-(1.2 \pm 0.6) \times 10^{-1}$ & $(2.8 \pm 0.7) \times 10^{-1}$ & $(8.6 \pm 6.0) \times 10^{-2}$ & $(3.3 \pm 0.7) \times 10^{-1}$ & $(1.3 \pm 0.6) \times 10^{-1}$ \\
$\xi_{\rsTR}$ & $-(1.2 \pm 0.6) \times 10^{-1}$ & $(8.9 \pm 5.8) \times 10^{-2}$ & $(-6.3 \pm 4.7) \times 10^{-2}$ & $(6.1 \pm 5.3) \times 10^{-2}$ & $(-6.6 \pm 4.3) \times 10^{-2}$ \\
$\varsigma_{\rsTL}$ & $(4.2 \pm 0.3) \times 10^{-1}$ & $(4.5 \pm 0.4) \times 10^{-1}$ & $(3.3 \pm 0.3) \times 10^{-1}$ & $(4.1 \pm 0.4) \times 10^{-1}$ & $(3.2 \pm 0.3) \times 10^{-1}$ \\
$\varsigma_{\rsTR}$ & $(4.2 \pm 0.3) \times 10^{-1}$ & $(4.1 \pm 0.3) \times 10^{-1}$ & $(3.3 \pm 0.2) \times 10^{-1}$ & $(4.4 \pm 0.3) \times 10^{-1}$ & $(3.4 \pm 0.2) \times 10^{-1}$ \\
\end{tabular}
\end{ruledtabular}
\end{table*}

If the log-returns $x_t$ are perfectly described by the fitted conditional volatility models, such that the normali\ENz{}ed residuals $z_t$ are i.i.d.\@ unit white noise, then exceedances of an arbitrary threshold should arrive as a Poisson point process. That is, the conditional intensity is not excited by past events ($\vect{\Gamma}=0$), and is instead equal to the constant background ($\vect{\uplambda}=\vect{\upmu}$). It would also be expected that the distribution of excess magnitudes $m_{k_\rsTO}$ would be stationary  and decay with $\mathrm{exp}{(-{m_{k_\rsTO}}^2)}$: the second of these conditions implies that the GPD shape parameter $\vect{\upxi}<0$. Finally, if the 2T-POT Hawkes model is fitted to i.i.d.\@ unit white noise using symmetrically defined exceedance thresholds $\vect{u}$, then all parameters should be symmetric, as is described by $\vect{\uptheta}\rsMdlSci$. 

\cref{fig:SP_cv_z_tt_G101_N_delta_t} shows that, without the leverage effect ($o=0$), the null hypothesis of i.i.d.\@ normali\ENz{}ed residuals is rejected at the 95\% confidence level with respect to right-tail exceedances, but not for left-tail exceedances; this result holds when the leverage effect is included ($o=1$, not shown). Thus, none of the four GARCH variants completely captures the data generating process. The fitted 2T-POT parameters in \cref{tab:params_cv_z_tt} show further deviations from the behavi\ENou{}r expected under the null. First, there is significant residual self-excitement of left-tail exceedances when $o=0$ and of right-tail exceedances when $o=1$\footnote{When $o=1$, $\beta_{\rsTR}^{-1}$ is found to be of the order $10^{3} \: \rsUdt$, meaning that the average daughter event occurs almost four years after the mother event. This does not describe clustering, instead the model is reflecting a long timescale variation that is not naturally described by a Hawkes-type self-exciting process.}, although the evidence for cross-excitement is much weaker. With the exception of the right-tail under Student's t errors, the fitted $\vect{\upxi}$ show that the tails of the GARCH residuals remain significantly heavy. All parameters but $\vect{\varsigma}$ exhibit significant asymmetries. Finally, we note that there is significant disagreement in the parameters across the different GARCH variants from which the residual series are derived. This recalls the point in \cref{sec:analysis}, namely that the residual exceedance events are highly sensitive to the assumptions of the chosen conditional volatility model, and, as a consequence, so is the fitting of a POT Hawkes model on the GARCH residuals.

\begin{table*}
\caption{
Fitted parameters ($\pm$ standard errors) for the 2T-POT Hawkes bivariate model $\vect{\uptheta}\rsMdlBi$ fitted to S\&P 500 daily log-returns $x_t | \mathrm{S\&P \: 500}$ and \num{50} simulated series of log-returns from the GARCH models $x_t | {\mathrm{G}{(0,\mathcal{N})}}$, $x_t | {\mathrm{G}{(0,t)}}$, $x_t | {\mathrm{G}{(1,\mathcal{N})}}$, and $x_t | {\mathrm{G}{(1,t)}}$.}
\label{tab:params_cv_x_tt}
\begin{ruledtabular}
\begin{tabular}{clllll}
Parameter & $x_t | \mathrm{S\&P \: 500}$ & $x_t | {\mathrm{G}{(0,\mathcal{N})}}$ & $x_t | {\mathrm{G}{(0,t)}}$ & $x_t | {\mathrm{G}{(1,\mathcal{N})}}$ & $x_t | {\mathrm{G}{(1,t)}}$\\
\hline
$u_{\rsTL}$ & $-1.840\times 10^{-2}$ & $(-2.4 \pm 0.3) \times 10^{-2}$ & $(-2.0 \pm 0.3) \times 10^{-2}$ & $(-2.0 \pm 0.7) \times 10^{-2}$ & $(-1.8 \pm 0.3) \times 10^{-2}$ \\
$u_{\rsTR}$ & $1.872\times 10^{-2}$ & $(2.4 \pm 0.3) \times 10^{-2}$ & $(2.1 \pm 0.3) \times 10^{-2}$ & $(2.1 \pm 0.8) \times 10^{-2}$ & $(1.8 \pm 0.2) \times 10^{-2}$ \\
$\mu_{\rsTL} / \rsUdt^{{-1}}$ & $(4.9 \pm 1.2) \times 10^{-3}$ & $(2.3 \pm 0.3) \times 10^{-2}$ & $(2.3 \pm 0.3) \times 10^{-2}$ & $(1.3 \pm 0.2) \times 10^{-2}$ & $(1.3 \pm 0.3) \times 10^{-2}$ \\
$\mu_{\rsTR} / \rsUdt^{{-1}}$ & $(3.1 \pm 0.8) \times 10^{-3}$ & $(2.2 \pm 0.3) \times 10^{-2}$ & $(2.2 \pm 0.3) \times 10^{-2}$ & $(3.2 \pm 0.3) \times 10^{-2}$ & $(3.3 \pm 0.3) \times 10^{-2}$ \\
$\gamma_{\rsTL\rsTL}$ & $(5.8 \pm 0.7) \times 10^{-1}$ & $(4.2 \pm 0.6) \times 10^{-1}$ & $(4.1 \pm 0.7) \times 10^{-1}$ & $(6.2 \pm 0.6) \times 10^{-1}$ & $(6.1 \pm 0.7) \times 10^{-1}$ \\
$\gamma_{\rsTL\rsTR}$ & $(2.2 \pm 0.8) \times 10^{-1}$ & $(4.6 \pm 0.7) \times 10^{-1}$ & $(4.3 \pm 0.6) \times 10^{-1}$ & $(2.0 \pm 0.5) \times 10^{-1}$ & $(1.9 \pm 0.7) \times 10^{-1}$ \\
$\gamma_{\rsTR\rsTL}$ & $(6.0 \pm 0.6) \times 10^{-1}$ & $(4.4 \pm 0.7) \times 10^{-1}$ & $(4.3 \pm 0.7) \times 10^{-1}$ & $(6.4 \pm 0.7) \times 10^{-1}$ & $(6.5 \pm 0.7) \times 10^{-1}$ \\
$\gamma_{\rsTR\rsTR}$ & $(2.8 \pm 0.6) \times 10^{-1}$ & $(4.3 \pm 0.7) \times 10^{-1}$ & $(4.1 \pm 0.6) \times 10^{-1}$ & $(1.9 \pm 0.5) \times 10^{-1}$ & $(1.6 \pm 0.6) \times 10^{-1}$ \\
$\beta_{\rsTL} / \rsUdt^{{-1}}$ & $(7.4 \pm 1.0) \times 10^{-2}$ & $(5.3 \pm 0.9) \times 10^{-2}$ & $(4.4 \pm 0.7) \times 10^{-2}$ & $(5.2 \pm 0.7) \times 10^{-2}$ & $(4.8 \pm 0.6) \times 10^{-2}$ \\
$\beta_{\rsTR} / \rsUdt^{{-1}}$ & $(1.7 \pm 0.4) \times 10^{-2}$ & $(5.0 \pm 0.9) \times 10^{-2}$ & $(4.4 \pm 0.8) \times 10^{-2}$ & $(5.5 \pm 1.4) \times 10^{-2}$ & $(4.7 \pm 1.5) \times 10^{-2}$ \\
$\xi_{\rsTL}$ & $(2.2 \pm 0.6) \times 10^{-1}$ & $(-1.5 \pm 0.6) \times 10^{-1}$ & $(1.3 \pm 6.2) \times 10^{-2}$ & $(-1.4 \pm 0.6) \times 10^{-1}$ & $(1.4 \pm 5.1) \times 10^{-2}$ \\
$\xi_{\rsTR}$ & $(-3.1 \pm 7.4) \times 10^{-2}$ & $(-1.3 \pm 0.6) \times 10^{-1}$ & $(1.9 \pm 5.9) \times 10^{-2}$ & $(-1.4 \pm 0.7) \times 10^{-1}$ & $(2.0 \pm 7.2) \times 10^{-2}$ \\
$\varsigma_{\rsTL}$ & $(3.8 \pm 0.5) \times 10^{-3}$ & $(3.8 \pm 0.7) \times 10^{-3}$ & $(4.6 \pm 0.9) \times 10^{-3}$ & $(3.6 \pm 1.2) \times 10^{-3}$ & $(4.1 \pm 0.7) \times 10^{-3}$ \\
$\varsigma_{\rsTR}$ & $(3.4 \pm 0.6) \times 10^{-3}$ & $(3.8 \pm 0.8) \times 10^{-3}$ & $(4.6 \pm 0.8) \times 10^{-3}$ & $(3.3 \pm 0.6) \times 10^{-3}$ & $(4.0 \pm 0.8) \times 10^{-3}$ \\
$\eta_{\rsTL}$ & $(3.2 \pm 0.9) \times 10^{-2}$ & $(7.7 \pm 1.9) \times 10^{-2}$ & $(6.3 \pm 1.5) \times 10^{-2}$ & $(7.0 \pm 7.1) \times 10^{-2}$ & $(5.4 \pm 1.7) \times 10^{-2}$ \\
$\eta_{\rsTR}$ & $(5.2 \pm 0.8) \times 10^{-2}$ & $(7.4 \pm 1.9) \times 10^{-2}$ & $(6.1 \pm 1.7) \times 10^{-2}$ & $(7.1 \pm 7.3) \times 10^{-2}$ & $(5.5 \pm 1.5) \times 10^{-2}$ \\
$\alpha_{\rsTL}$ & $(3.6 \pm 2.0) \times 10^{-1}$ & $1.3 \pm 1.7$ & $8.8 \pm 24.5$ & $1.1 \pm 1.6$ & $1.5 \pm 0.8$ \\
$\alpha_{\rsTR}$ & $2.2 \pm 3.6$ & $1.0 \pm 0.7$ & $9.0 \pm 24.2$ & $6.1 \pm 20.4$ & $(2.5 \pm 4.3) \times 10^{1}$ \\
\end{tabular}
\end{ruledtabular}
\end{table*}

We also use the fitted conditional volatility models to simulate log-returns. Using each GARCH model specified in \cref{tab:GARCH}, we generated \num{50} series of length equal to the original S\&P 500 training data (\num{12311}), then fit the 2T-POT Hawkes model to each of these simulated series. \cref{tab:params_cv_x_tt} compares the average parameter values for the simulated data against those  for the S\&P 500 daily log-returns. Without the leverage parameter ($o=0$) the conditional volatility models are fully symmetric with respect to the 2T-POT model (i.e.\ENLA{} they are naturally described by $\vect{\uptheta}\rsMdlSci$). The introduction of the leverage parameter ($o=1$) introduces asymmetries in $\vect{\upmu}$ and $\vect{\Gamma}$ that match with those seen in the original S\&P 500 data, meaning that the asymmetries in the background arrival rates and the branching matrix may be generated by the GJR-GARCH leverage effect. However, the asymmetries in the decay constant $\vect{\upbeta}$ and GPD tail shape parameter $\vect{\upxi}$ observed in the original data are not reproduced by any conditional volatility model: here, the 2T-POT model detects an aspect of the data generating process that cannot be attributed to the volatility clustering or the leverage effect as described by GARCH-type models. Instead, the asymmetry in $\vect{\upbeta}$ suggests a more complex version of the leverage effect, whereby the greater impact of losses on volatility is more pronounced over shorter timescales.

\section{\label{sec:ML}Maximum likelihood (ML) estimation}

\begin{figure*}[t]
\includegraphics{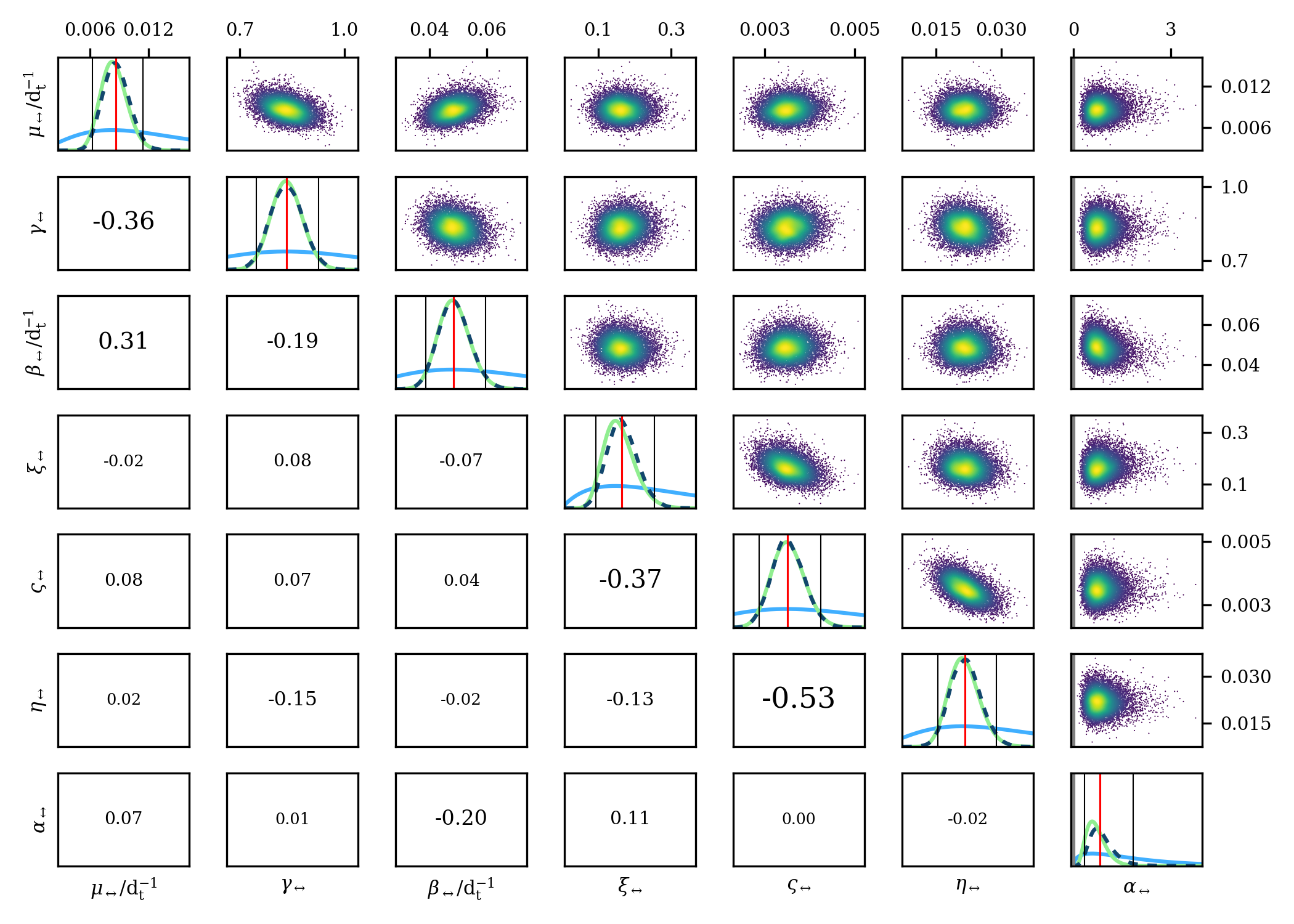}
\caption{
\label{fig:SP_MCMC_ci_s-mark}
Bayesian Markov chain Monte Carlo (MCMC) exploration of the $\vect{\uptheta}\rsMdlSci{}$ parameter space for extreme S\&P 500 daily log-returns. The upper-right off-diagonal panels show scatter plots of the trace in two-parameter sub-spaces; the lower-left off-diagonal panels show the sample correlations between these parameter pairs. The diagonal panels show the distributions of the L-BFGS-B parameter estimates and standard errors (light green), the Bayesian prior (blue), and a kernel density estimate of the trace (dark blue, dashed) along with its median (vertical, red) and 95\% confidence interval (vertical, black).
}
\end{figure*}

The parameters of the 2T-POT Hawkes models are found through maximum likelihood (ML) estimation. The log-likelihood under the parameters $\vect{\uptheta}$ over the data $\mathcal{I}_t$ is
\begin{widetext}
\begin{equation}
\label{eqn:logL}
\ell{\left(\vect{\uptheta} \middle| \mathcal{I}_t\right)} = 
\sum_{i}{
- \int_{0}^{t}{\lambda_{i}{\left(t' \middle|\bm{\uptheta}; \mathcal{I}_t\right)}d{t'}}
+ \sum_{k_i: t_{k_i}<t}{
\ln{\lambda_{i}{\left(t_{k_i} \middle|\bm{\uptheta}; \mathcal{I}_t\right)}} 
+ \ln{f_i{\left(m_{k_i} \middle| t_{k_i}\right)}}
}} ,
\end{equation}
\end{widetext}
where $i \in \{\rsTL, \rsTR\}$ for the bivariate model and $i \in \{\rsTA
\}$ for the common intensity model \cite{Gresnigt2015}.

We use the L-BFGS-B algorithm \cite{Byrd1995, Zhu1997} to find the parameters that maximi\ENz{}e \cref{eqn:logL}, 
\begin{equation}
\label{eqn:param_est}
\hat{\vect{\uptheta}}{\left(\mathcal{I}_t\right)} = \argmax_{\vect{\uptheta}}{\ell{\left(\vect{\uptheta} \middle| \mathcal{I}_t\right)}} .
\end{equation}
The standard errors of the parameter estimates, $\mathrm{SE}_{\hat{\vect{\uptheta}}}$, are then obtained by finite difference approximation of the Hessian matrix. The estimated $\hat{\vect{\uptheta}}$ and $\mathrm{SE}_{\hat{\vect{\uptheta}}}$ are shown in \cref{tab:params_mark_bi,tab:params_mark_ci}; the corresponding deviances, $-2\ell{(\hat{\vect{\uptheta}} | \mathcal{I}_t)}$, are given in \cref{tab:likelihood} under the training period columns.

We demonstrate the reliability of this procedure by comparing the estimated values and standard errors of the parameters against a Bayesian Markov chain Monte Carlo (MCMC) exploration of the parameter space using the No-U-Turn sampler in PyMC3 \cite{Homan2014}. As shown in \cref{fig:SP_MCMC_ci_s-mark}, the Bayesian posterior distributions clearly converge to the estimates and standard errors found via our numerical optimi\ENz{}ation procedure.

\bibliography{Bibliography/references}

\begin{thebibliography}{52}%
\makeatletter
\providecommand \@ifxundefined [1]{%
 \@ifx{#1\undefined}
}%
\providecommand \@ifnum [1]{%
 \ifnum #1\expandafter \@firstoftwo
 \else \expandafter \@secondoftwo
 \fi
}%
\providecommand \@ifx [1]{%
 \ifx #1\expandafter \@firstoftwo
 \else \expandafter \@secondoftwo
 \fi
}%
\providecommand \natexlab [1]{#1}%
\providecommand \enquote  [1]{``#1''}%
\providecommand \bibnamefont  [1]{#1}%
\providecommand \bibfnamefont [1]{#1}%
\providecommand \citenamefont [1]{#1}%
\providecommand \href@noop [0]{\@secondoftwo}%
\providecommand \href [0]{\begingroup \@sanitize@url \@href}%
\providecommand \@href[1]{\@@startlink{#1}\@@href}%
\providecommand \@@href[1]{\endgroup#1\@@endlink}%
\providecommand \@sanitize@url [0]{\catcode `\\12\catcode `\$12\catcode
  `\&12\catcode `\#12\catcode `\^12\catcode `\_12\catcode `\%12\relax}%
\providecommand \@@startlink[1]{}%
\providecommand \@@endlink[0]{}%
\providecommand \url  [0]{\begingroup\@sanitize@url \@url }%
\providecommand \@url [1]{\endgroup\@href {#1}{\urlprefix }}%
\providecommand \urlprefix  [0]{URL }%
\providecommand \Eprint [0]{\href }%
\providecommand \doibase [0]{https://doi.org/}%
\providecommand \selectlanguage [0]{\@gobble}%
\providecommand \bibinfo  [0]{\@secondoftwo}%
\providecommand \bibfield  [0]{\@secondoftwo}%
\providecommand \translation [1]{[#1]}%
\providecommand \BibitemOpen [0]{}%
\providecommand \bibitemStop [0]{}%
\providecommand \bibitemNoStop [0]{.\EOS\space}%
\providecommand \EOS [0]{\spacefactor3000\relax}%
\providecommand \BibitemShut  [1]{\csname bibitem#1\endcsname}%
\let\auto@bib@innerbib\@empty
\bibitem [{\citenamefont {Kim}\ \emph {et~al.}(2020)\citenamefont {Kim},
  \citenamefont {Paini},\ and\ \citenamefont {Jurdak}}]{Kim2020}%
  \BibitemOpen
  \bibfield  {author} {\bibinfo {author} {\bibfnamefont {M.}~\bibnamefont
  {Kim}}, \bibinfo {author} {\bibfnamefont {D.}~\bibnamefont {Paini}},\ and\
  \bibinfo {author} {\bibfnamefont {R.}~\bibnamefont {Jurdak}},\ }\bibfield
  {title} {\bibinfo {title} {{Real-world diffusion dynamics based on point
  process approaches: a review}},\ }\href
  {https://doi.org/10.1007/s10462-018-9656-9} {\bibfield  {journal} {\bibinfo
  {journal} {Artif. Intell. Rev.}\ }\textbf {\bibinfo {volume} {53}},\ \bibinfo
  {pages} {321} (\bibinfo {year} {2020})}\BibitemShut {NoStop}%
\bibitem [{\citenamefont {Reinhart}(2018)}]{Reinhart2018}%
  \BibitemOpen
  \bibfield  {author} {\bibinfo {author} {\bibfnamefont {A.}~\bibnamefont
  {Reinhart}},\ }\bibfield  {title} {\bibinfo {title} {{A Review of
  Self-Exciting Spatio-Temporal Point Processes and Their Applications}},\
  }\href {https://doi.org/10.1214/17-STS629} {\bibfield  {journal} {\bibinfo
  {journal} {Stat. Sci.}\ }\textbf {\bibinfo {volume} {33}},\ \bibinfo {pages}
  {299} (\bibinfo {year} {2018})}\BibitemShut {NoStop}%
\bibitem [{\citenamefont {Coles}(2001)}]{Coles2001}%
  \BibitemOpen
  \bibfield  {author} {\bibinfo {author} {\bibfnamefont {S.}~\bibnamefont
  {Coles}},\ }\href {https://doi.org/10.1007/978-1-4471-3675-0} {\emph
  {\bibinfo {title} {{An Introduction to Statistical Modeling of Extreme
  Values}}}},\ \bibinfo {series} {Springer Series in Statistics}, Vol.\
  \bibinfo {volume} {208}\ (\bibinfo  {publisher} {Springer London},\ \bibinfo
  {address} {London},\ \bibinfo {year} {2001})\BibitemShut {NoStop}%
\bibitem [{\citenamefont {de~Haan}\ and\ \citenamefont
  {Ferreira}(2006)}]{DeHaan2006}%
  \BibitemOpen
  \bibfield  {author} {\bibinfo {author} {\bibfnamefont {L.}~\bibnamefont
  {de~Haan}}\ and\ \bibinfo {author} {\bibfnamefont {A.}~\bibnamefont
  {Ferreira}},\ }\href {https://doi.org/10.1007/0-387-34471-3} {\emph {\bibinfo
  {title} {{Extreme Value Theory}}}},\ Springer Series in Operations Research
  and Financial Engineering\ (\bibinfo  {publisher} {Springer New York},\
  \bibinfo {address} {New York},\ \bibinfo {year} {2006})\BibitemShut {NoStop}%
\bibitem [{\citenamefont {Scarrott}\ and\ \citenamefont
  {MacDonald}(2012)}]{Scarrott2012}%
  \BibitemOpen
  \bibfield  {author} {\bibinfo {author} {\bibfnamefont {C.}~\bibnamefont
  {Scarrott}}\ and\ \bibinfo {author} {\bibfnamefont {A.}~\bibnamefont
  {MacDonald}},\ }\bibfield  {title} {\bibinfo {title} {{A review of extreme
  value threshold estimation and uncertainty quantification}},\ }\href
  {https://www.ine.pt/revstat/pdf/rs120102.pdf} {\bibfield  {journal} {\bibinfo
   {journal} {Revstat-Stat. J.}\ }\textbf {\bibinfo {volume} {10}},\ \bibinfo
  {pages} {33} (\bibinfo {year} {2012})}\BibitemShut {NoStop}%
\bibitem [{\citenamefont {Hawkes}(1971{\natexlab{a}})}]{Hawkes1971a}%
  \BibitemOpen
  \bibfield  {author} {\bibinfo {author} {\bibfnamefont {A.~G.}\ \bibnamefont
  {Hawkes}},\ }\bibfield  {title} {\bibinfo {title} {{Spectra of some
  self-exciting and mutually exciting point processes}},\ }\href
  {https://doi.org/10.2307/2334319} {\bibfield  {journal} {\bibinfo  {journal}
  {Biometrika}\ }\textbf {\bibinfo {volume} {58}},\ \bibinfo {pages} {83}
  (\bibinfo {year} {1971}{\natexlab{a}})}\BibitemShut {NoStop}%
\bibitem [{\citenamefont {Hawkes}(1971{\natexlab{b}})}]{Hawkes1971b}%
  \BibitemOpen
  \bibfield  {author} {\bibinfo {author} {\bibfnamefont {A.~G.}\ \bibnamefont
  {Hawkes}},\ }\bibfield  {title} {\bibinfo {title} {{Point Spectra of Some
  Mutually Exciting Point Processes}},\ }\href
  {https://doi.org/10.1111/j.2517-6161.1971.tb01530.x} {\bibfield  {journal}
  {\bibinfo  {journal} {J. R. Stat. Soc. Ser. B (Methodol.)}\ }\textbf
  {\bibinfo {volume} {33}},\ \bibinfo {pages} {438} (\bibinfo {year}
  {1971}{\natexlab{b}})}\BibitemShut {NoStop}%
\bibitem [{\citenamefont {Hawkes}(2018)}]{Hawkes2018}%
  \BibitemOpen
  \bibfield  {author} {\bibinfo {author} {\bibfnamefont {A.~G.}\ \bibnamefont
  {Hawkes}},\ }\bibfield  {title} {\bibinfo {title} {{Hawkes processes and
  their applications to finance: a review}},\ }\href
  {https://doi.org/10.1080/14697688.2017.1403131} {\bibfield  {journal}
  {\bibinfo  {journal} {Quant. Financ.}\ }\textbf {\bibinfo {volume} {18}},\
  \bibinfo {pages} {193} (\bibinfo {year} {2018})}\BibitemShut {NoStop}%
\bibitem [{\citenamefont {Adamopoulos}(1976)}]{Adamopoulos1976}%
  \BibitemOpen
  \bibfield  {author} {\bibinfo {author} {\bibfnamefont {L.}~\bibnamefont
  {Adamopoulos}},\ }\bibfield  {title} {\bibinfo {title} {{Cluster models for
  earthquakes: Regional comparisons}},\ }\href
  {https://doi.org/10.1007/BF01028982} {\bibfield  {journal} {\bibinfo
  {journal} {J. Int. Assoc. Math. Geol.}\ }\textbf {\bibinfo {volume} {8}},\
  \bibinfo {pages} {463} (\bibinfo {year} {1976})}\BibitemShut {NoStop}%
\bibitem [{\citenamefont {Shcherbakov}\ \emph {et~al.}(2019)\citenamefont
  {Shcherbakov}, \citenamefont {Zhuang}, \citenamefont {Z{\"{o}}ller},\ and\
  \citenamefont {Ogata}}]{Shcherbakov2019}%
  \BibitemOpen
  \bibfield  {author} {\bibinfo {author} {\bibfnamefont {R.}~\bibnamefont
  {Shcherbakov}}, \bibinfo {author} {\bibfnamefont {J.}~\bibnamefont {Zhuang}},
  \bibinfo {author} {\bibfnamefont {G.}~\bibnamefont {Z{\"{o}}ller}},\ and\
  \bibinfo {author} {\bibfnamefont {Y.}~\bibnamefont {Ogata}},\ }\bibfield
  {title} {\bibinfo {title} {{Forecasting the magnitude of the largest expected
  earthquake}},\ }\href {https://doi.org/10.1038/s41467-019-11958-4} {\bibfield
   {journal} {\bibinfo  {journal} {Nat. Commun.}\ }\textbf {\bibinfo {volume}
  {10}},\ \bibinfo {pages} {4051} (\bibinfo {year} {2019})}\BibitemShut
  {NoStop}%
\bibitem [{\citenamefont {Pernice}\ \emph {et~al.}(2012)\citenamefont
  {Pernice}, \citenamefont {Staude}, \citenamefont {Cardanobile},\ and\
  \citenamefont {Rotter}}]{Pernice2012}%
  \BibitemOpen
  \bibfield  {author} {\bibinfo {author} {\bibfnamefont {V.}~\bibnamefont
  {Pernice}}, \bibinfo {author} {\bibfnamefont {B.}~\bibnamefont {Staude}},
  \bibinfo {author} {\bibfnamefont {S.}~\bibnamefont {Cardanobile}},\ and\
  \bibinfo {author} {\bibfnamefont {S.}~\bibnamefont {Rotter}},\ }\bibfield
  {title} {\bibinfo {title} {{Recurrent interactions in spiking networks with
  arbitrary topology}},\ }\href {https://doi.org/10.1103/PhysRevE.85.031916}
  {\bibfield  {journal} {\bibinfo  {journal} {Phys. Rev. E}\ }\textbf {\bibinfo
  {volume} {85}},\ \bibinfo {pages} {031916} (\bibinfo {year}
  {2012})}\BibitemShut {NoStop}%
\bibitem [{\citenamefont {Tannenbaum}\ and\ \citenamefont
  {Burak}(2017)}]{Tannenbaum2017}%
  \BibitemOpen
  \bibfield  {author} {\bibinfo {author} {\bibfnamefont {N.~R.}\ \bibnamefont
  {Tannenbaum}}\ and\ \bibinfo {author} {\bibfnamefont {Y.}~\bibnamefont
  {Burak}},\ }\bibfield  {title} {\bibinfo {title} {{Theory of nonstationary
  Hawkes processes}},\ }\href {https://doi.org/10.1103/PhysRevE.96.062314}
  {\bibfield  {journal} {\bibinfo  {journal} {Phys. Rev. E}\ }\textbf {\bibinfo
  {volume} {96}},\ \bibinfo {pages} {062314} (\bibinfo {year}
  {2017})}\BibitemShut {NoStop}%
\bibitem [{\citenamefont {Short}\ \emph {et~al.}(2014)\citenamefont {Short},
  \citenamefont {Mohler}, \citenamefont {Brantingham},\ and\ \citenamefont
  {Tita}}]{Short2014}%
  \BibitemOpen
  \bibfield  {author} {\bibinfo {author} {\bibfnamefont {M.~B.}\ \bibnamefont
  {Short}}, \bibinfo {author} {\bibfnamefont {G.~O.}\ \bibnamefont {Mohler}},
  \bibinfo {author} {\bibfnamefont {P.~J.}\ \bibnamefont {Brantingham}},\ and\
  \bibinfo {author} {\bibfnamefont {G.~E.}\ \bibnamefont {Tita}},\ }\bibfield
  {title} {\bibinfo {title} {{Gang rivalry dynamics via coupled point process
  networks}},\ }\href {https://doi.org/10.3934/dcdsb.2014.19.1459} {\bibfield
  {journal} {\bibinfo  {journal} {Discrete Contin. Dyn.-B}\ }\textbf {\bibinfo
  {volume} {19}},\ \bibinfo {pages} {1459} (\bibinfo {year}
  {2014})}\BibitemShut {NoStop}%
\bibitem [{\citenamefont {Johnson}\ \emph {et~al.}(2018)\citenamefont
  {Johnson}, \citenamefont {Hitchman}, \citenamefont {Phan},\ and\
  \citenamefont {Smith}}]{Johnson2018}%
  \BibitemOpen
  \bibfield  {author} {\bibinfo {author} {\bibfnamefont {N.}~\bibnamefont
  {Johnson}}, \bibinfo {author} {\bibfnamefont {A.}~\bibnamefont {Hitchman}},
  \bibinfo {author} {\bibfnamefont {D.}~\bibnamefont {Phan}},\ and\ \bibinfo
  {author} {\bibfnamefont {L.}~\bibnamefont {Smith}},\ }\bibfield  {title}
  {\bibinfo {title} {{Self-exciting point process models for political conflict
  forecasting}},\ }\href {https://doi.org/10.1017/S095679251700033X} {\bibfield
   {journal} {\bibinfo  {journal} {Eur. J. Appl. Math.}\ }\textbf {\bibinfo
  {volume} {29}},\ \bibinfo {pages} {685} (\bibinfo {year} {2018})}\BibitemShut
  {NoStop}%
\bibitem [{\citenamefont {Fujita}\ \emph {et~al.}(2018)\citenamefont {Fujita},
  \citenamefont {Medvedev}, \citenamefont {Koyama}, \citenamefont {Lambiotte},\
  and\ \citenamefont {Shinomoto}}]{Fujita2018}%
  \BibitemOpen
  \bibfield  {author} {\bibinfo {author} {\bibfnamefont {K.}~\bibnamefont
  {Fujita}}, \bibinfo {author} {\bibfnamefont {A.}~\bibnamefont {Medvedev}},
  \bibinfo {author} {\bibfnamefont {S.}~\bibnamefont {Koyama}}, \bibinfo
  {author} {\bibfnamefont {R.}~\bibnamefont {Lambiotte}},\ and\ \bibinfo
  {author} {\bibfnamefont {S.}~\bibnamefont {Shinomoto}},\ }\bibfield  {title}
  {\bibinfo {title} {{Identifying exogenous and endogenous activity in social
  media}},\ }\href {https://doi.org/10.1103/PhysRevE.98.052304} {\bibfield
  {journal} {\bibinfo  {journal} {Phys. Rev. E}\ }\textbf {\bibinfo {volume}
  {98}},\ \bibinfo {pages} {052304} (\bibinfo {year} {2018})}\BibitemShut
  {NoStop}%
\bibitem [{\citenamefont {Bacry}\ \emph {et~al.}(2015)\citenamefont {Bacry},
  \citenamefont {Mastromatteo},\ and\ \citenamefont {Muzy}}]{Bacry2015}%
  \BibitemOpen
  \bibfield  {author} {\bibinfo {author} {\bibfnamefont {E.}~\bibnamefont
  {Bacry}}, \bibinfo {author} {\bibfnamefont {I.}~\bibnamefont
  {Mastromatteo}},\ and\ \bibinfo {author} {\bibfnamefont {J.-F.}\ \bibnamefont
  {Muzy}},\ }\bibfield  {title} {\bibinfo {title} {{Hawkes processes in
  finance}},\ }\href {https://doi.org/10.1142/S2382626615500057} {\bibfield
  {journal} {\bibinfo  {journal} {Mark. Microstruct. Liq.}\ }\textbf {\bibinfo
  {volume} {1}},\ \bibinfo {pages} {1550005} (\bibinfo {year}
  {2015})}\BibitemShut {NoStop}%
\bibitem [{\citenamefont {Hawkes}(2020)}]{Hawkes2020}%
  \BibitemOpen
  \bibfield  {author} {\bibinfo {author} {\bibfnamefont {A.~G.}\ \bibnamefont
  {Hawkes}},\ }\bibfield  {title} {\bibinfo {title} {{Hawkes jump-diffusions
  and finance: a brief history and review}},\ }\href
  {https://doi.org/10.1080/1351847X.2020.1755712} {\bibfield  {journal}
  {\bibinfo  {journal} {Eur. J. Financ.}\ }\textbf {\bibinfo {volume} {0}},\
  \bibinfo {pages} {1} (\bibinfo {year} {2020})}\BibitemShut {NoStop}%
\bibitem [{\citenamefont {Bowsher}(2007)}]{Bowsher2007}%
  \BibitemOpen
  \bibfield  {author} {\bibinfo {author} {\bibfnamefont {C.~G.}\ \bibnamefont
  {Bowsher}},\ }\bibfield  {title} {\bibinfo {title} {{Modelling security
  market events in continuous time: Intensity based, multivariate point process
  models}},\ }\href
  {https://doi.org/https://doi.org/10.1016/j.jeconom.2006.11.007} {\bibfield
  {journal} {\bibinfo  {journal} {J. Econom.}\ }\textbf {\bibinfo {volume}
  {141}},\ \bibinfo {pages} {876} (\bibinfo {year} {2007})}\BibitemShut
  {NoStop}%
\bibitem [{\citenamefont {Bacry}\ \emph {et~al.}(2012)\citenamefont {Bacry},
  \citenamefont {Dayri},\ and\ \citenamefont {Muzy}}]{Bacry2012}%
  \BibitemOpen
  \bibfield  {author} {\bibinfo {author} {\bibfnamefont {E.}~\bibnamefont
  {Bacry}}, \bibinfo {author} {\bibfnamefont {K.}~\bibnamefont {Dayri}},\ and\
  \bibinfo {author} {\bibfnamefont {J.~F.}\ \bibnamefont {Muzy}},\ }\bibfield
  {title} {\bibinfo {title} {{Non-parametric kernel estimation for symmetric
  Hawkes processes. Application to high frequency financial data}},\ }\href
  {https://doi.org/10.1140/epjb/e2012-21005-8} {\bibfield  {journal} {\bibinfo
  {journal} {Eur. Phys. J. B}\ }\textbf {\bibinfo {volume} {85}},\ \bibinfo
  {pages} {157} (\bibinfo {year} {2012})}\BibitemShut {NoStop}%
\bibitem [{\citenamefont {Filimonov}\ and\ \citenamefont
  {Sornette}(2012)}]{Filimonov2012}%
  \BibitemOpen
  \bibfield  {author} {\bibinfo {author} {\bibfnamefont {V.}~\bibnamefont
  {Filimonov}}\ and\ \bibinfo {author} {\bibfnamefont {D.}~\bibnamefont
  {Sornette}},\ }\bibfield  {title} {\bibinfo {title} {{Quantifying reflexivity
  in financial markets: Toward a prediction of flash crashes}},\ }\href
  {https://doi.org/10.1103/PhysRevE.85.056108} {\bibfield  {journal} {\bibinfo
  {journal} {Phys. Rev. E}\ }\textbf {\bibinfo {volume} {85}},\ \bibinfo
  {pages} {056108} (\bibinfo {year} {2012})}\BibitemShut {NoStop}%
\bibitem [{\citenamefont {Hardiman}\ \emph {et~al.}(2013)\citenamefont
  {Hardiman}, \citenamefont {Bercot},\ and\ \citenamefont
  {Bouchaud}}]{Hardiman2013}%
  \BibitemOpen
  \bibfield  {author} {\bibinfo {author} {\bibfnamefont {S.~J.}\ \bibnamefont
  {Hardiman}}, \bibinfo {author} {\bibfnamefont {N.}~\bibnamefont {Bercot}},\
  and\ \bibinfo {author} {\bibfnamefont {J.-P.}\ \bibnamefont {Bouchaud}},\
  }\bibfield  {title} {\bibinfo {title} {{Critical reflexivity in financial
  markets: a Hawkes process analysis}},\ }\href
  {https://doi.org/10.1140/epjb/e2013-40107-3} {\bibfield  {journal} {\bibinfo
  {journal} {Eur. Phys. J. B}\ }\textbf {\bibinfo {volume} {86}},\ \bibinfo
  {pages} {442} (\bibinfo {year} {2013})}\BibitemShut {NoStop}%
\bibitem [{\citenamefont {Hardiman}\ and\ \citenamefont
  {Bouchaud}(2014)}]{Hardiman2014}%
  \BibitemOpen
  \bibfield  {author} {\bibinfo {author} {\bibfnamefont {S.~J.}\ \bibnamefont
  {Hardiman}}\ and\ \bibinfo {author} {\bibfnamefont {J.-P.}\ \bibnamefont
  {Bouchaud}},\ }\bibfield  {title} {\bibinfo {title} {{Branching-ratio
  approximation for the self-exciting Hawkes process}},\ }\href
  {https://doi.org/10.1103/PhysRevE.90.062807} {\bibfield  {journal} {\bibinfo
  {journal} {Phys. Rev. E}\ }\textbf {\bibinfo {volume} {90}},\ \bibinfo
  {pages} {062807} (\bibinfo {year} {2014})}\BibitemShut {NoStop}%
\bibitem [{\citenamefont {Rambaldi}\ \emph {et~al.}(2015)\citenamefont
  {Rambaldi}, \citenamefont {Pennesi},\ and\ \citenamefont
  {Lillo}}]{Rambaldi2015}%
  \BibitemOpen
  \bibfield  {author} {\bibinfo {author} {\bibfnamefont {M.}~\bibnamefont
  {Rambaldi}}, \bibinfo {author} {\bibfnamefont {P.}~\bibnamefont {Pennesi}},\
  and\ \bibinfo {author} {\bibfnamefont {F.}~\bibnamefont {Lillo}},\ }\bibfield
   {title} {\bibinfo {title} {{Modeling foreign exchange market activity around
  macroeconomic news: Hawkes-process approach}},\ }\href
  {https://doi.org/10.1103/PhysRevE.91.012819} {\bibfield  {journal} {\bibinfo
  {journal} {Phys. Rev. E}\ }\textbf {\bibinfo {volume} {91}},\ \bibinfo
  {pages} {012819} (\bibinfo {year} {2015})}\BibitemShut {NoStop}%
\bibitem [{\citenamefont {Grothe}\ \emph {et~al.}(2014)\citenamefont {Grothe},
  \citenamefont {Korniichuk},\ and\ \citenamefont {Manner}}]{Grothe2014}%
  \BibitemOpen
  \bibfield  {author} {\bibinfo {author} {\bibfnamefont {O.}~\bibnamefont
  {Grothe}}, \bibinfo {author} {\bibfnamefont {V.}~\bibnamefont {Korniichuk}},\
  and\ \bibinfo {author} {\bibfnamefont {H.}~\bibnamefont {Manner}},\
  }\bibfield  {title} {\bibinfo {title} {{Modeling multivariate extreme events
  using self-exciting point processes}},\ }\href
  {https://doi.org/https://doi.org/10.1016/j.jeconom.2014.03.011} {\bibfield
  {journal} {\bibinfo  {journal} {J. Econom.}\ }\textbf {\bibinfo {volume}
  {182}},\ \bibinfo {pages} {269} (\bibinfo {year} {2014})}\BibitemShut
  {NoStop}%
\bibitem [{\citenamefont {Gresnigt}\ \emph {et~al.}(2015)\citenamefont
  {Gresnigt}, \citenamefont {Kole},\ and\ \citenamefont
  {Franses}}]{Gresnigt2015}%
  \BibitemOpen
  \bibfield  {author} {\bibinfo {author} {\bibfnamefont {F.}~\bibnamefont
  {Gresnigt}}, \bibinfo {author} {\bibfnamefont {E.}~\bibnamefont {Kole}},\
  and\ \bibinfo {author} {\bibfnamefont {P.~H.}\ \bibnamefont {Franses}},\
  }\bibfield  {title} {\bibinfo {title} {{Interpreting financial market crashes
  as earthquakes: A new Early Warning System for medium term crashes}},\ }\href
  {https://doi.org/10.1016/j.jbankfin.2015.03.003} {\bibfield  {journal}
  {\bibinfo  {journal} {J. Bank. Financ.}\ }\textbf {\bibinfo {volume} {56}},\
  \bibinfo {pages} {123} (\bibinfo {year} {2015})}\BibitemShut {NoStop}%
\bibitem [{\citenamefont {Gresnigt}\ \emph {et~al.}(2016)\citenamefont
  {Gresnigt}, \citenamefont {Kole},\ and\ \citenamefont
  {Franses}}]{Gresnigt2016}%
  \BibitemOpen
  \bibfield  {author} {\bibinfo {author} {\bibfnamefont {F.}~\bibnamefont
  {Gresnigt}}, \bibinfo {author} {\bibfnamefont {E.}~\bibnamefont {Kole}},\
  and\ \bibinfo {author} {\bibfnamefont {P.~H.}\ \bibnamefont {Franses}},\
  }\bibfield  {title} {\bibinfo {title} {{Specification Testing in Hawkes
  Models}},\ }\href {https://doi.org/10.1093/jjfinec/nbw011} {\bibfield
  {journal} {\bibinfo  {journal} {J. Financ. Econom.}\ }\textbf {\bibinfo
  {volume} {15}},\ \bibinfo {pages} {139} (\bibinfo {year} {2016})}\BibitemShut
  {NoStop}%
\bibitem [{\citenamefont {Gresnigt}\ \emph {et~al.}(2017)\citenamefont
  {Gresnigt}, \citenamefont {Kole},\ and\ \citenamefont
  {Franses}}]{Gresnigt2017}%
  \BibitemOpen
  \bibfield  {author} {\bibinfo {author} {\bibfnamefont {F.}~\bibnamefont
  {Gresnigt}}, \bibinfo {author} {\bibfnamefont {E.}~\bibnamefont {Kole}},\
  and\ \bibinfo {author} {\bibfnamefont {P.~H.}\ \bibnamefont {Franses}},\
  }\bibfield  {title} {\bibinfo {title} {{Exploiting Spillovers to Forecast
  Crashes}},\ }\href {https://doi.org/10.1002/for.2434} {\bibfield  {journal}
  {\bibinfo  {journal} {J. Forecast.}\ }\textbf {\bibinfo {volume} {36}},\
  \bibinfo {pages} {936} (\bibinfo {year} {2017})}\BibitemShut {NoStop}%
\bibitem [{\citenamefont {Chavez-Demoulin}\ \emph {et~al.}(2005)\citenamefont
  {Chavez-Demoulin}, \citenamefont {Davison},\ and\ \citenamefont
  {McNeil}}]{Chavez-Demoulin2005}%
  \BibitemOpen
  \bibfield  {author} {\bibinfo {author} {\bibfnamefont {V.}~\bibnamefont
  {Chavez-Demoulin}}, \bibinfo {author} {\bibfnamefont {A.~C.}\ \bibnamefont
  {Davison}},\ and\ \bibinfo {author} {\bibfnamefont {A.~J.}\ \bibnamefont
  {McNeil}},\ }\bibfield  {title} {\bibinfo {title} {{Estimating value-at-risk:
  a point process approach}},\ }\href
  {https://doi.org/10.1080/14697680500039613} {\bibfield  {journal} {\bibinfo
  {journal} {Quant. Financ.}\ }\textbf {\bibinfo {volume} {5}},\ \bibinfo
  {pages} {227} (\bibinfo {year} {2005})}\BibitemShut {NoStop}%
\bibitem [{\citenamefont {Chavez-Demoulin}\ and\ \citenamefont
  {McGill}(2012)}]{Chavez-Demoulin2012}%
  \BibitemOpen
  \bibfield  {author} {\bibinfo {author} {\bibfnamefont {V.}~\bibnamefont
  {Chavez-Demoulin}}\ and\ \bibinfo {author} {\bibfnamefont {J.~A.}\
  \bibnamefont {McGill}},\ }\bibfield  {title} {\bibinfo {title}
  {{High-frequency financial data modeling using Hawkes processes}},\ }\href
  {https://doi.org/https://doi.org/10.1016/j.jbankfin.2012.08.011} {\bibfield
  {journal} {\bibinfo  {journal} {J. Bank. Financ.}\ }\textbf {\bibinfo
  {volume} {36}},\ \bibinfo {pages} {3415} (\bibinfo {year}
  {2012})}\BibitemShut {NoStop}%
\bibitem [{\citenamefont {Bie{\'{n}}-Barkowska}(2020)}]{BienBarkowska2020}%
  \BibitemOpen
  \bibfield  {author} {\bibinfo {author} {\bibfnamefont {K.}~\bibnamefont
  {Bie{\'{n}}-Barkowska}},\ }\bibfield  {title} {\bibinfo {title} {{Looking at
  Extremes without Going to Extremes: A New Self-Exciting Probability Model for
  Extreme Losses in Financial Markets}},\ }\href
  {https://doi.org/10.3390/e22070789} {\bibfield  {journal} {\bibinfo
  {journal} {Entropy}\ }\textbf {\bibinfo {volume} {22}},\ \bibinfo {pages}
  {789} (\bibinfo {year} {2020})}\BibitemShut {NoStop}%
\bibitem [{\citenamefont {Baumeister}\ \emph {et~al.}(2001)\citenamefont
  {Baumeister}, \citenamefont {Bratslavsky}, \citenamefont {Finkenauer},\ and\
  \citenamefont {Vohs}}]{Baumeister2001}%
  \BibitemOpen
  \bibfield  {author} {\bibinfo {author} {\bibfnamefont {R.~F.}\ \bibnamefont
  {Baumeister}}, \bibinfo {author} {\bibfnamefont {E.}~\bibnamefont
  {Bratslavsky}}, \bibinfo {author} {\bibfnamefont {C.}~\bibnamefont
  {Finkenauer}},\ and\ \bibinfo {author} {\bibfnamefont {K.~D.}\ \bibnamefont
  {Vohs}},\ }\bibfield  {title} {\bibinfo {title} {{Bad is Stronger than
  Good}},\ }\href {https://doi.org/10.1037/1089-2680.5.4.323} {\bibfield
  {journal} {\bibinfo  {journal} {Rev. Gen. Psychol.}\ }\textbf {\bibinfo
  {volume} {5}},\ \bibinfo {pages} {323} (\bibinfo {year} {2001})}\BibitemShut
  {NoStop}%
\bibitem [{\citenamefont {Rozin}\ and\ \citenamefont
  {Royzman}(2001)}]{Rozin2001}%
  \BibitemOpen
  \bibfield  {author} {\bibinfo {author} {\bibfnamefont {P.}~\bibnamefont
  {Rozin}}\ and\ \bibinfo {author} {\bibfnamefont {E.~B.}\ \bibnamefont
  {Royzman}},\ }\bibfield  {title} {\bibinfo {title} {{Negativity Bias,
  Negativity Dominance, and Contagion}},\ }\href
  {https://doi.org/10.1207/S15327957PSPR0504_2} {\bibfield  {journal} {\bibinfo
   {journal} {Personal. Soc. Psychol. Rev.}\ }\textbf {\bibinfo {volume} {5}},\
  \bibinfo {pages} {296} (\bibinfo {year} {2001})}\BibitemShut {NoStop}%
\bibitem [{\citenamefont {Bachelier}(1900)}]{Bachelier1900}%
  \BibitemOpen
  \bibfield  {author} {\bibinfo {author} {\bibfnamefont {L.}~\bibnamefont
  {Bachelier}},\ }\bibfield  {title} {\bibinfo {title} {{Th{\'{e}}orie de la
  sp{\'{e}}culation}},\ }\href {https://doi.org/10.24033/asens.476} {\bibfield
  {journal} {\bibinfo  {journal} {Ann. Sci. \'{E}cole Norm. S.}\ }\textbf
  {\bibinfo {volume} {17}},\ \bibinfo {pages} {21} (\bibinfo {year}
  {1900})}\BibitemShut {NoStop}%
\bibitem [{\citenamefont {Ruppert}\ and\ \citenamefont
  {Matteson}(2015)}]{Ruppert2015a}%
  \BibitemOpen
  \bibfield  {author} {\bibinfo {author} {\bibfnamefont {D.}~\bibnamefont
  {Ruppert}}\ and\ \bibinfo {author} {\bibfnamefont {D.~S.}\ \bibnamefont
  {Matteson}},\ }\href {https://doi.org/10.1007/978-1-4939-2614-5} {\emph
  {\bibinfo {title} {{Statistics and Data Analysis for Financial
  Engineering}}}},\ \bibinfo {edition} {2nd}\ ed.,\ Springer Texts in
  Statistics\ (\bibinfo  {publisher} {Springer New York},\ \bibinfo {address}
  {New York},\ \bibinfo {year} {2015})\BibitemShut {NoStop}%
\bibitem [{\citenamefont {Cont}(2001)}]{Cont2001}%
  \BibitemOpen
  \bibfield  {author} {\bibinfo {author} {\bibfnamefont {R.}~\bibnamefont
  {Cont}},\ }\bibfield  {title} {\bibinfo {title} {{Empirical properties of
  asset returns: Stylized facts and statistical issues}},\ }\href
  {https://doi.org/10.1080/713665670} {\bibfield  {journal} {\bibinfo
  {journal} {Quant. Financ.}\ }\textbf {\bibinfo {volume} {1}},\ \bibinfo
  {pages} {223} (\bibinfo {year} {2001})}\BibitemShut {NoStop}%
\bibitem [{\citenamefont {Davies}\ and\ \citenamefont
  {Kr{\"{a}}mer}(2016)}]{Davies2015}%
  \BibitemOpen
  \bibfield  {author} {\bibinfo {author} {\bibfnamefont {L.}~\bibnamefont
  {Davies}}\ and\ \bibinfo {author} {\bibfnamefont {W.}~\bibnamefont
  {Kr{\"{a}}mer}},\ }\bibfield  {title} {\bibinfo {title} {{Stylized Facts and
  Simulating Long Range Financial Data}},\ }\href
  {http://arxiv.org/abs/1612.05229} {\  (\bibinfo {year} {2016})},\ \Eprint
  {https://arxiv.org/abs/1612.05229} {arXiv:1612.05229} \BibitemShut {NoStop}%
\bibitem [{\citenamefont {Tsay}(2010)}]{Tsay2010}%
  \BibitemOpen
  \bibfield  {author} {\bibinfo {author} {\bibfnamefont {R.~S.}\ \bibnamefont
  {Tsay}},\ }\href {https://doi.org/10.1002/9780470644560} {\emph {\bibinfo
  {title} {{Analysis of Financial Time Series}}}},\ \bibinfo {edition} {3rd}\
  ed.,\ Wiley Series in Probability and Statistics\ (\bibinfo  {publisher}
  {John Wiley {\&} Sons, Inc.},\ \bibinfo {address} {Hoboken, NJ},\ \bibinfo
  {year} {2010})\BibitemShut {NoStop}%
\bibitem [{\citenamefont {Lopes}(2011)}]{Lopes2011}%
  \BibitemOpen
  \bibfield  {author} {\bibinfo {author} {\bibfnamefont {R.~H.~C.}\
  \bibnamefont {Lopes}},\ }\bibinfo {title} {{Kolmogorov-Smirnov Test}},\ in\
  \href {https://doi.org/10.1007/978-3-642-04898-2_326} {\emph {\bibinfo
  {booktitle} {International Encyclopedia of Statistical Science}}},\ \bibinfo
  {editor} {edited by\ \bibinfo {editor} {\bibfnamefont {M.}~\bibnamefont
  {Lovric}}}\ (\bibinfo  {publisher} {Springer Berlin Heidelberg},\ \bibinfo
  {address} {Berlin, Heidelberg},\ \bibinfo {year} {2011})\ pp.\ \bibinfo
  {pages} {718--720}\BibitemShut {NoStop}%
\bibitem [{\citenamefont {Embrechts}\ \emph {et~al.}(2011)\citenamefont
  {Embrechts}, \citenamefont {Liniger},\ and\ \citenamefont
  {Lin}}]{Embrechts2011}%
  \BibitemOpen
  \bibfield  {author} {\bibinfo {author} {\bibfnamefont {P.}~\bibnamefont
  {Embrechts}}, \bibinfo {author} {\bibfnamefont {T.}~\bibnamefont {Liniger}},\
  and\ \bibinfo {author} {\bibfnamefont {L.}~\bibnamefont {Lin}},\ }\bibfield
  {title} {\bibinfo {title} {{Multivariate Hawkes processes: an application to
  financial data}},\ }\href {https://doi.org/10.1017/S0021900200099344}
  {\bibfield  {journal} {\bibinfo  {journal} {J. Appl. Probab.}\ }\textbf
  {\bibinfo {volume} {48}},\ \bibinfo {pages} {367} (\bibinfo {year}
  {2011})}\BibitemShut {NoStop}%
\bibitem [{\citenamefont {Chicheportiche}\ and\ \citenamefont
  {Chakraborti}(2014)}]{Chicheportiche2014}%
  \BibitemOpen
  \bibfield  {author} {\bibinfo {author} {\bibfnamefont {R.}~\bibnamefont
  {Chicheportiche}}\ and\ \bibinfo {author} {\bibfnamefont {A.}~\bibnamefont
  {Chakraborti}},\ }\bibfield  {title} {\bibinfo {title} {{Copulas and time
  series with long-ranged dependencies}},\ }\href
  {https://doi.org/10.1103/PhysRevE.89.042117} {\bibfield  {journal} {\bibinfo
  {journal} {Phys. Rev. E}\ }\textbf {\bibinfo {volume} {89}},\ \bibinfo
  {pages} {042117} (\bibinfo {year} {2014})}\BibitemShut {NoStop}%
\bibitem [{\citenamefont {Rabin}(1998)}]{Rabin1998}%
  \BibitemOpen
  \bibfield  {author} {\bibinfo {author} {\bibfnamefont {M.}~\bibnamefont
  {Rabin}},\ }\bibfield  {title} {\bibinfo {title} {{Psychology and
  Economics}},\ }\href {http://www.jstor.org/stable/2564950} {\bibfield
  {journal} {\bibinfo  {journal} {J. Econ. Lit.}\ }\textbf {\bibinfo {volume}
  {36}},\ \bibinfo {pages} {11} (\bibinfo {year} {1998})}\BibitemShut {NoStop}%
\bibitem [{\citenamefont {Kahneman}(2003)}]{Khaneman2003}%
  \BibitemOpen
  \bibfield  {author} {\bibinfo {author} {\bibfnamefont {D.}~\bibnamefont
  {Kahneman}},\ }\bibfield  {title} {\bibinfo {title} {{Maps of bounded
  rationality: Psychology for behavioral economics}},\ }\href
  {https://doi.org/10.1257/000282803322655392} {\bibfield  {journal} {\bibinfo
  {journal} {Am. Econ. Rev.}\ }\textbf {\bibinfo {volume} {93}},\ \bibinfo
  {pages} {1449} (\bibinfo {year} {2003})}\BibitemShut {NoStop}%
\bibitem [{\citenamefont {Pickands}(1975)}]{Pickands1975}%
  \BibitemOpen
  \bibfield  {author} {\bibinfo {author} {\bibfnamefont {J.}~\bibnamefont
  {Pickands}},\ }\bibfield  {title} {\bibinfo {title} {{Statistical Inference
  Using Extreme Order Statistics}},\ }\href
  {http://www.jstor.org/stable/2958083} {\bibfield  {journal} {\bibinfo
  {journal} {Ann. Stat.}\ }\textbf {\bibinfo {volume} {3}},\ \bibinfo {pages}
  {119} (\bibinfo {year} {1975})}\BibitemShut {NoStop}%
\bibitem [{\citenamefont {Balkema}\ and\ \citenamefont
  {de~Haan}(1974)}]{Balkema1974}%
  \BibitemOpen
  \bibfield  {author} {\bibinfo {author} {\bibfnamefont {A.~A.}\ \bibnamefont
  {Balkema}}\ and\ \bibinfo {author} {\bibfnamefont {L.}~\bibnamefont
  {de~Haan}},\ }\bibfield  {title} {\bibinfo {title} {{Residual Life Time at
  Great Age}},\ }\href {http://www.jstor.org/stable/2959306} {\bibfield
  {journal} {\bibinfo  {journal} {Ann. Probab.}\ }\textbf {\bibinfo {volume}
  {2}},\ \bibinfo {pages} {792} (\bibinfo {year} {1974})}\BibitemShut {NoStop}%
\bibitem [{\citenamefont {Wheatley}\ \emph {et~al.}(2019)\citenamefont
  {Wheatley}, \citenamefont {Wehrli},\ and\ \citenamefont
  {Sornette}}]{Wheatley2019}%
  \BibitemOpen
  \bibfield  {author} {\bibinfo {author} {\bibfnamefont {S.}~\bibnamefont
  {Wheatley}}, \bibinfo {author} {\bibfnamefont {A.}~\bibnamefont {Wehrli}},\
  and\ \bibinfo {author} {\bibfnamefont {D.}~\bibnamefont {Sornette}},\
  }\bibfield  {title} {\bibinfo {title} {{The endo–exo problem in high
  frequency financial price fluctuations and rejecting criticality}},\ }\href
  {https://doi.org/10.1080/14697688.2018.1550266} {\bibfield  {journal}
  {\bibinfo  {journal} {Quant. Financ.}\ }\textbf {\bibinfo {volume} {19}},\
  \bibinfo {pages} {1165} (\bibinfo {year} {2019})}\BibitemShut {NoStop}%
\bibitem [{Yah(2020)}]{YahooFinance2020}%
  \BibitemOpen
  \href {https://finance.yahoo.com/} {\bibinfo {title} {{Yahoo Finance}}}
  (\bibinfo {year} {2020})\BibitemShut {NoStop}%
\bibitem [{\citenamefont {Glosten}\ \emph {et~al.}(1993)\citenamefont
  {Glosten}, \citenamefont {Jagannathan},\ and\ \citenamefont
  {Runkle}}]{Glosten1993}%
  \BibitemOpen
  \bibfield  {author} {\bibinfo {author} {\bibfnamefont {L.~R.}\ \bibnamefont
  {Glosten}}, \bibinfo {author} {\bibfnamefont {R.}~\bibnamefont
  {Jagannathan}},\ and\ \bibinfo {author} {\bibfnamefont {D.~E.}\ \bibnamefont
  {Runkle}},\ }\bibfield  {title} {\bibinfo {title} {{On the Relation between
  the Expected Value and the Volatility of the Nominal Excess Return on
  Stocks}},\ }\href {https://doi.org/10.1111/j.1540-6261.1993.tb05128.x}
  {\bibfield  {journal} {\bibinfo  {journal} {J. Finance}\ }\textbf {\bibinfo
  {volume} {48}},\ \bibinfo {pages} {1779} (\bibinfo {year}
  {1993})}\BibitemShut {NoStop}%
\bibitem [{\citenamefont {Wit}\ \emph {et~al.}(2012)\citenamefont {Wit},
  \citenamefont {van~den Heuvel},\ and\ \citenamefont {Romeijn}}]{Wit2012}%
  \BibitemOpen
  \bibfield  {author} {\bibinfo {author} {\bibfnamefont {E.}~\bibnamefont
  {Wit}}, \bibinfo {author} {\bibfnamefont {E.}~\bibnamefont {van~den
  Heuvel}},\ and\ \bibinfo {author} {\bibfnamefont {J.-W.}\ \bibnamefont
  {Romeijn}},\ }\bibfield  {title} {\bibinfo {title} {{‘All models are
  wrong...': an introduction to model uncertainty}},\ }\href
  {https://doi.org/10.1111/j.1467-9574.2012.00530.x} {\bibfield  {journal}
  {\bibinfo  {journal} {Stat. Neerl.}\ }\textbf {\bibinfo {volume} {66}},\
  \bibinfo {pages} {217} (\bibinfo {year} {2012})}\BibitemShut {NoStop}%
\bibitem [{\citenamefont {Ogata}(1988)}]{Ogata1988}%
  \BibitemOpen
  \bibfield  {author} {\bibinfo {author} {\bibfnamefont {Y.}~\bibnamefont
  {Ogata}},\ }\bibfield  {title} {\bibinfo {title} {{Statistical Models for
  Earthquake Occurrences and Residual Analysis for Point Processes}},\ }\href
  {https://doi.org/10.1080/01621459.1988.10478560} {\bibfield  {journal}
  {\bibinfo  {journal} {J. Am. Stat. Assoc.}\ }\textbf {\bibinfo {volume}
  {83}},\ \bibinfo {pages} {9} (\bibinfo {year} {1988})}\BibitemShut {NoStop}%
\bibitem [{\citenamefont {Byrd}\ \emph {et~al.}(1995)\citenamefont {Byrd},
  \citenamefont {Lu}, \citenamefont {Nocedal},\ and\ \citenamefont
  {Zhu}}]{Byrd1995}%
  \BibitemOpen
  \bibfield  {author} {\bibinfo {author} {\bibfnamefont {R.~H.}\ \bibnamefont
  {Byrd}}, \bibinfo {author} {\bibfnamefont {P.}~\bibnamefont {Lu}}, \bibinfo
  {author} {\bibfnamefont {J.}~\bibnamefont {Nocedal}},\ and\ \bibinfo {author}
  {\bibfnamefont {C.}~\bibnamefont {Zhu}},\ }\bibfield  {title} {\bibinfo
  {title} {{A Limited Memory Algorithm for Bound Constrained Optimization}},\
  }\href {https://doi.org/10.1137/0916069} {\bibfield  {journal} {\bibinfo
  {journal} {SIAM J. Sci. Comput.}\ }\textbf {\bibinfo {volume} {16}},\
  \bibinfo {pages} {1190} (\bibinfo {year} {1995})}\BibitemShut {NoStop}%
\bibitem [{\citenamefont {Zhu}\ \emph {et~al.}(1997)\citenamefont {Zhu},
  \citenamefont {Byrd}, \citenamefont {Lu},\ and\ \citenamefont
  {Nocedal}}]{Zhu1997}%
  \BibitemOpen
  \bibfield  {author} {\bibinfo {author} {\bibfnamefont {C.}~\bibnamefont
  {Zhu}}, \bibinfo {author} {\bibfnamefont {R.~H.}\ \bibnamefont {Byrd}},
  \bibinfo {author} {\bibfnamefont {P.}~\bibnamefont {Lu}},\ and\ \bibinfo
  {author} {\bibfnamefont {J.}~\bibnamefont {Nocedal}},\ }\bibfield  {title}
  {\bibinfo {title} {{Algorithm 778: L-BFGS-B: Fortran Subroutines for
  Large-Scale Bound-Constrained Optimization}},\ }\href
  {https://doi.org/10.1145/279232.279236} {\bibfield  {journal} {\bibinfo
  {journal} {ACM Trans. Math. Softw.}\ }\textbf {\bibinfo {volume} {23}},\
  \bibinfo {pages} {550} (\bibinfo {year} {1997})}\BibitemShut {NoStop}%
\bibitem [{\citenamefont {Homan}\ and\ \citenamefont
  {Gelman}(2014)}]{Homan2014}%
  \BibitemOpen
  \bibfield  {author} {\bibinfo {author} {\bibfnamefont {M.~D.}\ \bibnamefont
  {Homan}}\ and\ \bibinfo {author} {\bibfnamefont {A.}~\bibnamefont {Gelman}},\
  }\bibfield  {title} {\bibinfo {title} {{The No-U-Turn Sampler: Adaptively
  Setting Path Lengths in Hamiltonian Monte Carlo}},\ }\href
  {https://dl.acm.org/doi/10.5555/2627435.2638586} {\bibfield  {journal}
  {\bibinfo  {journal} {J. Mach. Learn. Res.}\ }\textbf {\bibinfo {volume}
  {15}},\ \bibinfo {pages} {1593} (\bibinfo {year} {2014})}\BibitemShut
  {NoStop}%
\end{thebibliography}%

\end{document}